\documentclass[a4paper,12pt]{article}
\pdfoutput=1 

\usepackage{jheppub} 
\usepackage[T1]{fontenc} 
\usepackage[utf8]{inputenc}
\usepackage{tikz}

\usepackage{caption}
\usepackage{subcaption}
\usepackage{shuffle}
\usepackage{amssymb,amsmath}
\usepackage{bbm}
\usepackage{bm}
\usepackage{enumitem}
\usepackage{mathtools}
\usepackage{mathrsfs}
\usepackage{graphicx}
\usepackage{epsfig}
\usepackage{epstopdf}
\usepackage{setspace}
\usepackage{dsfont}
\usepackage{latexsym}
\usepackage{tikz}
\usepackage{psfrag}
\usepackage{booktabs}
\usepackage{bbm}
\usepackage[toc]{appendix}
\usepackage{color}
\usepackage{datetime}
\usepackage{tensor}
\usepackage{lscape}
\usepackage{comment}
\usepackage{lipsum}

\usetikzlibrary{arrows,shapes}
\usetikzlibrary{trees}
\usetikzlibrary{patterns}
\usetikzlibrary{matrix,arrows} 				
\usetikzlibrary{positioning}				
\usetikzlibrary{calc,through}				
\usetikzlibrary{decorations.pathreplacing}  
\usepackage{pgffor}							
\usetikzlibrary{decorations}
\usetikzlibrary{decorations.pathmorphing}	
\usetikzlibrary{decorations.markings}
\usetikzlibrary{intersections}
\tikzset{
    vector/.style={decorate, decoration={snake}, draw},
	provector/.style={decorate, decoration={snake,amplitude=2.5pt}, draw},
	antivector/.style={decorate, decoration={snake,amplitude=-2.5pt}, draw},
        smallvector/.style={decorate, decoration={snake,amplitude=1.5pt,post length=0.5mm}, draw},
    fermion/.style={draw=black, postaction={decorate},
        decoration={markings,mark=at position .55 with {\arrow[draw=black]{>}}}},
    fermionbar/.style={draw=black, postaction={decorate},
        decoration={markings,mark=at position .55 with {\arrow[draw=black]{<}}}},
    fermionnoarrow/.style={draw=black},
    gluon/.style={decorate, draw=black,
        decoration={coil,amplitude=4pt, segment length=5pt}},
    scalar/.style={dashed,draw=black, postaction={decorate},
        decoration={markings,mark=at position .55 with {\arrow[draw=black]{>}}}},
    scalarbar/.style={dashed,draw=black, postaction={decorate},
        decoration={markings,mark=at position .55 with {\arrow[draw=black]{<}}}},
    scalarnoarrow/.style={dashed,draw=black},
    electron/.style={draw=black, postaction={decorate},
        decoration={markings,mark=at position .55 with {\arrow[draw=black]{>}}}},
    bigvector/.style={decorate, decoration={snake,amplitude=4pt}, draw},
    arrow/.style={draw=black, postaction={decorate},
        decoration={markings,mark=at position 1 with {\arrow[draw=black]{>}}}},
	sexchange/.pic={\tikzset{every node/.style={font=\scriptsize}}
    \pgfmathsetmacro{\r}{0.75}
		\draw [] (0,0) circle (\r cm);
		\tikzset{decoration={snake,amplitude=.4mm,segment length=1.5mm,post length=0mm,pre length=0mm}}
		\filldraw (45:\r) circle (1pt) node[above right=-4pt]{$\Delta_3$};
		\filldraw (135:\r) circle (1pt) node[above left=-4pt]{$\Delta_2$};
		\filldraw (-135:\r) circle (1pt) node[below left=-4pt]{$\Delta_1$};
		\filldraw (-45:\r) circle (1pt) node[below right=-4pt]{$\Delta_4$};
		\filldraw (-0.3,0) circle (1pt) (0.3,0) circle (1pt);
		\draw [thick] (-0.3,0) -- (0.3,0);
		\draw [thick] (45:\r) -- (0.3,0);
		\draw [thick] (-45:\r) -- (0.3,0);
		\draw [thick] (135:\r) -- (-0.3,0);
		\draw [thick] (-135:\r) -- (-0.3,0);
	\node at (0,0) [below=0pt]{$\Delta$};
	},
	sf/.pic={\tikzset{every node/.style={font=\scriptsize}}
    \pgfmathsetmacro{\r}{0.75}
		\draw [] (0,0) circle (\r cm);
		\tikzset{decoration={snake,amplitude=.4mm,segment length=1.5mm,post length=0mm,pre length=0mm}}
		\filldraw (-120:\r) circle (1pt) node[label={[below=4pt]$1$}] (a1) {};
		\filldraw (180:\r) circle (1pt) node[label={[yshift=-0.35cm,xshift=-5pt]$2$}] (a2) {};
		\filldraw (120:\r) circle (1pt) node[label={[above=-3pt]$3$}] (a3) {};
		\filldraw (60:\r) circle (1pt) node[label={[above=-3pt]$4$}] (a4) {};
		\filldraw (0:\r) circle (1pt) node[label={[yshift=-0.35cm,xshift=5pt]$5$}] (a5) {};
		\filldraw (-60:\r) circle (1pt) node[label={[below=4pt]$6$}] (a6) {};
		\filldraw (-150:2*\r/3) circle (1pt) coordinate (v1);
		\filldraw (90:2*\r/3) circle (1pt) coordinate (v2);
		\filldraw (-30:2*\r/3) circle (1pt) coordinate (v3);
		\filldraw (0,0) circle (1pt) coordinate (o);
		\draw [thick] (a1.center) -- (v1) -- (a2.center);
		\draw [thick] (a3.center) -- (v2) -- (a4.center);
		\draw [thick] (a5.center) -- (v3) -- (a6.center);
		\draw [thick,decorate] (v1) -- (o);
		\draw [thick,decorate] (v2) -- (o);
		\draw [thick,decorate] (v3) -- (o);
	},
	hl/.pic={\tikzset{every node/.style={font=\scriptsize}}
    \pgfmathsetmacro{\r}{0.75}
		\draw [] (0,0) circle (\r cm);
		\tikzset{decoration={snake,amplitude=.4mm,segment length=1.5mm,post length=0mm,pre length=0mm}}
		\filldraw (-120:\r) circle (1pt) node[label={[below=4pt]$1$}] (a1) {};
		\filldraw (180:\r) circle (1pt) node[label={[yshift=-0.35cm,xshift=-5pt]$2$}] (a2) {};
		\filldraw (120:\r) circle (1pt) node[label={[above=-3pt]$3$}] (a3) {};
		\filldraw (60:\r) circle (1pt) node[label={[above=-3pt]$4$}] (a4) {};
		\filldraw (0:\r) circle (1pt) node[label={[yshift=-0.35cm,xshift=5pt]$5$}] (a5) {};
		\filldraw (-60:\r) circle (1pt) node[label={[below=4pt]$6$}] (a6) {};
		\filldraw (-150:2*\r/3) circle (1pt) coordinate (v1);
		\filldraw (-30:2*\r/3) circle (1pt) coordinate (v2);
		\filldraw (120:\r/3) circle (1pt) coordinate (v3);
		\filldraw (60:\r/3) circle (1pt) coordinate (v4);
		\draw [thick] (a1.center) -- (v1) -- (a2.center);
		\draw [thick] (a5.center) -- (v2) -- (a6.center);
		\draw [thick] (a3.center) -- (v3) -- (v4) -- (a4.center);
		\draw [thick,decorate] (v1) -- (v3);
		\draw [thick,decorate] (v2) -- (v4);
	},
	ddmhl/.pic={\tikzset{every node/.style={font=\scriptsize}}
    \pgfmathsetmacro{\r}{0.75}
		\draw [] (0,0) circle (\r cm);
		\draw [thick] (-\r,0) -- (\r,0);
		\filldraw (-\r,0) circle (1pt) node[label={[yshift=-0.35cm,xshift=-5pt]$1$}] (a1) {};
		\filldraw (\r,0) circle (1pt) node[label={[yshift=-0.35cm,xshift=5pt]$6$}] (a6) {};
		\filldraw ({acos(0.2)}:\r) circle (1pt) node[label={[above=-5pt]$4$}] (a4) {};
		\filldraw ({acos(0.6)}:\r) circle (1pt) node[label={[above=-5pt]$5$}] (a5) {};
        \filldraw ({acos(-0.2)}:\r) circle (1pt) node[label={[above=-5pt]$3$}] (a3) {};
		\filldraw ({acos(-0.6)}:\r) circle (1pt) node[label={[above=-5pt]$2$}] (a2) {};
		\filldraw [thick] (a2.center) -- (-3*\r/5,0) circle (1pt);
		\filldraw [thick] (a3.center) -- (-\r/5,0) circle (1pt);
		\filldraw [thick] (a4.center) -- (\r/5,0) circle (1pt);
		\filldraw [thick] (a5.center) -- (3*\r/5,0) circle (1pt);
		\node[label={[above=-3pt]$\phantom{4}$}] at (60:\r) {}; 
		\node[label={[below=4pt]$\phantom{6}$}] at (-60:\r) {}; 
	}
}

\tikzstyle{block} = [draw, rectangle, 
    minimum height=3em, minimum width=6earticlem]

\definecolor{darkblue}{rgb}{0.2, 0, 0.8}

\numberwithin{equation}{section}

\renewcommand{\r}{\rho}

\newcommand{\prop}{\mathcal{G}}
\newcommand{\Cdot}{\!\cdot\!}
\newcommand{\qc}[1]{\mathsf{M}^2_{#1}}
\newcommand{\DD}{\mathcal{D}}

\newcommand{\gexchange}[5]{
\begin{tikzpicture}[baseline={([yshift=#1]current bounding box.center)},every node/.style={font=\scriptsize}]
\pgfmathsetmacro{\r}{0.75}
			\draw [] (0,0) circle (\r cm);
			\tikzset{decoration={snake,amplitude=.4mm,segment length=1.25mm,post length=0mm,pre length=0mm}}
			\filldraw (45:\r) circle (1pt) node[above=0pt]{$#4$};
			\filldraw (135:\r) circle (1pt) node[above=0pt]{$#3$};
			\filldraw (-135:\r) circle (1pt) node[below=0pt]{$#2$};
			\filldraw (-45:\r) circle (1pt) node[below=0pt]{$#5$};
			\filldraw (-0.3,0) circle (1pt) (0.3,0) circle (1pt);
			\draw [thick,decorate] (-0.3,0) -- (0.3,0);
			\draw [thick] (45:\r) -- (0.3,0);
			\draw [thick] (-45:\r) -- (0.3,0);
			\draw [thick] (135:\r) -- (-0.3,0);
			\draw [thick] (-135:\r) -- (-0.3,0);
\end{tikzpicture}
}
\newcommand{\snowflake}{
\begin{tikzpicture}[scale=0.25]
    \pgfmathsetmacro{\r}{0.75}
		\coordinate (a1) at (-120:\r);
		\coordinate (a2) at (180:\r);
		\coordinate (a3) at (120:\r);
		\coordinate (a4) at (60:\r);
		\coordinate (a5) at (0:\r);
		\coordinate (a6) at (-60:\r);
		\coordinate (v1) at (-150:2*\r/3);
		\coordinate (v2) at (90:2*\r/3);
		\coordinate (v3) at (-30:2*\r/3);
		\coordinate (o) at (0,0);
		\draw (a1) -- (v1) -- (a2);
		\draw (a3) -- (v2) -- (a4);
		\draw (a5) -- (v3) -- (a6);
		\draw (v1) -- (o);
		\draw (v2) -- (o);
		\draw (v3) -- (o);
\end{tikzpicture}
}
\newcommand{\cont}{
\begin{tikzpicture}[scale=0.25]
    \pgfmathsetmacro{\r}{0.75}
		\coordinate (a1) at (-120:\r);
		\coordinate (a2) at (180:\r);
		\coordinate (a3) at (120:\r);
		\coordinate (a4) at (60:\r);
		\coordinate (a5) at (0:\r);
		\coordinate (a6) at (-60:\r);
		\coordinate (v1) at (-150:2*\r/3);
		\coordinate (v2) at (90:2*\r/3);
		\coordinate (v3) at (-30:\r/3);
		\draw (a1) -- (v1) -- (a2);
		\draw (a3) -- (v2) -- (a4);
		\draw (a5) -- (v3) -- (a6);
		\draw (v1) -- (v3);
		\draw (v2) -- (v3);
\end{tikzpicture}
}
\newcommand{\halfladder}{
\begin{tikzpicture}[scale=0.4]
    \draw (0,0) -- (1.25,0);
    \foreach \x in {0.25,0.5,0.75,1}
    \draw (\x,0) -- ++(0,0.25);
\end{tikzpicture}
}
\newcommand{\hexchange}[5]{
\begin{tikzpicture}[baseline={([yshift=#1]current bounding box.center)},every node/.style={font=\scriptsize}]
\pgfmathsetmacro{\r}{0.75}
			\draw [] (0,0) circle (\r cm);
			\tikzset{decoration={snake,amplitude=.4mm,segment length=1.25mm,post length=0mm,pre length=0mm}}
			\filldraw (45:\r) circle (1pt) node[above=0pt]{$#4$};
			\filldraw (135:\r) circle (1pt) node[above=0pt]{$#3$};
			\filldraw (-135:\r) circle (1pt) node[below=0pt]{$#2$};
			\filldraw (-45:\r) circle (1pt) node[below=0pt]{$#5$};
			\filldraw (-0.3,0) circle (1pt) (0.3,0) circle (1pt);
			\draw [thick,double,decorate] (-0.3,0) -- (0.3,0);
			\draw [thick] (45:\r) -- (0.3,0);
			\draw [thick] (-45:\r) -- (0.3,0);
			\draw [thick] (135:\r) -- (-0.3,0);
			\draw [thick] (-135:\r) -- (-0.3,0);
\end{tikzpicture}
}
\newcommand{\ctd}[7]{
\begin{tikzpicture}[baseline={([yshift=#1]current bounding box.center)},every node/.style={font=\scriptsize}]
\pgfmathsetmacro{\r}{0.75}
		\draw [] (0,0) circle (\r cm);
		\tikzset{decoration={snake,amplitude=.4mm,segment length=1.5mm,post length=0mm,pre length=0mm}}
		\filldraw (-120:\r) circle (1pt) node[label={[below=4pt]$#2$}] (a1) {};
		\filldraw (180:\r) circle (1pt) node[label={[yshift=-0.35cm,xshift=-5pt]$#3$}] (a2) {};
		\filldraw (120:\r) circle (1pt) node[label={[above=-3pt]$#4$}] (a3) {};
		\filldraw (60:\r) circle (1pt) node[label={[above=-3pt]$#5$}] (a4) {};
		\filldraw (0:\r) circle (1pt) node[label={[yshift=-0.35cm,xshift=5pt]$#6$}] (a5) {};
		\filldraw (-60:\r) circle (1pt) node[label={[below=4pt]$#7$}] (a6) {};
		\filldraw (-150:\r/2) circle (1pt) coordinate (v1);
		\filldraw (90:\r/2) circle (1pt) coordinate (v2);
		\filldraw (-30:\r/2) circle (1pt) coordinate (v3);
		\draw [thick] (a1.center) -- (v1) -- (a2.center);
		\draw [thick] (a3.center) -- (v2) -- (a4.center);
		\draw [thick] (a5.center) -- (v3) -- (a6.center);
		\draw [thick,decorate] (v1) -- (v3);
		\draw [thick,decorate] (v2) -- (v3);
\end{tikzpicture}
}

\tikzset{vertex/.style={inner sep=0,minimum size=3pt,circle,fill}}

\usepackage{comment}

\usepackage{overpic}

\def\be{\begin{equation}}
\def\ee{\end{equation}}
\def\bea{\begin{eqnarray}}
\def\eea{\end{eqnarray}}
\def\ba{\begin{array}}
\def\ea{\end{array}}
\def\bd{\begin{displaymath}}
\def\ed{\end{displaymath}}

\def\r{\rho}                                     

\def\x{\xi}

\def\>{\rangle} 
\def\<{\langle} 
\def\Dsl{D \hskip-.6em \raise1pt\hbox{$ / $ } }

\def\ads{\textrm{AdS}}

\title{On the Differential Representation and Color-Kinematics Duality of AdS Boundary Correlators}

\author[a]{Aidan Herderschee,}
\author[b]{Radu Roiban}
\author[b]{and Fei Teng}

\affiliation[a]{Leinweber Center for Theoretical Physics, Randall Laboratory of Physics\\ The University of Michigan, Ann Arbor, MI 48109-1040, USA}
\affiliation[b]{Department of Physics, Pennsylvania State University, University Park, PA 16802, USA}

\emailAdd{aidanh@umich.edu}
\emailAdd{radu@phys.psu.edu}
\emailAdd{fei.teng@psu.edu}

\abstract{
The AdS boundary correlators and their dual correlation functions of boundary operators have been the main dynamic observables of the holographic duality relating a bulk AdS theory and a boundary conformal field theory.
We show that tree-level AdS boundary correlators for generic states can be expressed as nonlocal differential operators of a certain structure acting on contact Witten diagrams. We further write the boundary correlators in a form that is very similar to flat space amplitudes, with Mandelstam variables replaced by  certain combinations of single-state conformal generators, prove that all tree-level AdS boundary correlators have a differential representation, and detail the conversion of such differential expressions to position space.
We illustrate the construction through the computation of the boundary correlators of scalars coupled to gluons and gravitons; when converted to position space, they reproduce known results. 
Color-kinematics duality and BCJ relations can be defined in analogy with their flat space counterparts, and are respected by the scalar correlators with a gluon exchange. 
We also discuss potential approaches to the double copy and find that its direct generalization may require nontrivial extensions.
}

\preprint{LCTP-22-01}

\begin{document} 

\maketitle
\flushbottom



\section{Introduction}
\label{sec:intro}

The holographic duality~\cite{Susskind:1994vu,tHooft:1993dmi}, and its specific incarnation in which it relates field theories in anti-de Sitter (AdS) space and (possibly exotic) lines of conformal fixed points (CFT) on its boundary \cite{Maldacena:1997re, Gubser:1998bc, Witten:1998qj}, continues to be a source of insight into aspects of theories on both sides of the duality.
Boundary correlation functions in the bulk are the most natural observables to exploit the duality as they correspond to correlation functions of local CFT operators. 
They access (non-perturbative) quantum gravity in a curved space-time through strongly-coupled boundary calculations and reveal aspects of flat space CFTs at large coupling through weakly-coupled bulk calculations.

Weakly-coupled AdS boundary correlators are traditionally computed in terms of Witten diagrams, based on vertices derived from the bulk Lagrangian. The difficulty of such calculations in position space \cite{Arutyunov:2000py, Arutyunov:2002ff, Arutyunov:2002fh, DHoker:1999mqo, DHoker:1999kzh} is similar in spirit (although less severe because of the large symmetry of AdS space) with difficulties with tree-level calculations in general smooth curved spaces: even the simplest diagrams are given by nontrivial bulk integrals which lead to transcendental functions. This led to the development of several  alternative methods:  the AdS momentum space \cite{Raju:2012zr,Bzowski:2013sza,Farrow:2018yni,Albayrak:2018tam,Albayrak:2019yve,Bzowski:2019kwd,Lipstein:2019mpu,Albayrak:2020isk,Armstrong:2020woi,Albayrak:2020fyp,Jain:2021qcl,Jain:2021vrv}, position-Mellin space \cite{Mack:2009gy,Mack:2009mi,Penedones:2010ue,Paulos:2011ie,Fitzpatrick:2011ia,Kharel:2013mka,Penedones:2019tng,Rastelli:2016nze,Zhou:2017zaw,Rastelli:2017udc,Alday:2020dtb,Alday:2020lbp,Zhou:2021gnu,Alday:2021ajh,Alday:2022lkk} and momentum-Mellin space~\cite{Sleight:2021iix, Sleight:2019hfp}. 
While simplifying various aspects of calculations, these representations do not completely eliminate the problem of nontrivial integrals which still appear at tree-level in the momentum space formulation of AdS Witten diagrams and in the inverse Mellin-transform to position or momentum space. Moreover, even though spinning intermediate states and the relevant factorization have been describe in Mellin space \cite{Goncalves:2014rfa}, spinning external states have been more difficult to include. With notable exceptions~\cite{Sleight:2021iix,Sleight:2018epi}, computations of higher-spin correlators in Mellin space often rely on supersymmetry to relate them to scalar correlators, cf. e.g.~\cite{Rastelli:2016nze,Zhou:2017zaw,Rastelli:2017udc,Alday:2020dtb,Alday:2020lbp,Zhou:2021gnu,Alday:2021ajh}.

Other approaches to tree-level AdS boundary correlators, which avoid some of the previous technical difficulties and are inspired by the close analogy between holographic correlators and flat space scattering amplitudes, are the 
position-space and Mellin space analytic bootstraps introduced in~\cite{Rastelli:2016nze,Rastelli:2017udc,Goncalves:2019znr}. They evaluate $\ads_5$ boundary correlators in the supergravity limit in sufficiently-symmetric theories bypassing use of the Lagrangian, relying instead only on crossing symmetry, analyticity, superconformal symmetry, and the flat space limit. 
For less symmetric situations further information is required and may be obtained either from the bulk Lagrangian or from the expected dual CFT following the general discussion in \cite{Heemskerk:2009pn}.

Beginning with case studies, the differential representation has emerged as a new framework for studying AdS boundary correlators. It conjecturally expresses them as a nonlocal combination of generators of the AdS symmetry group at various boundary  points acting on a single contact integral determined by the mass parameters of the chosen fields.\footnote{Contact integrals or diagrams are defined in eq.~\eqref{contactint} below. As will become clear below, by ``nonlocal combination of generators'' we refer specifically to the appearance of inverses of scalar combinations of generators in these combinations.} It thus naturally manifests all symmetries, encompasses all of the various representations (position, momentum and Mellin), and may expose structures hidden under these traditional representations, such as the existence~\cite{Diwakar:2021juk} of differential relations between color-ordered boundary correlators analogous to the flat space Bern-Carrasco-Johansson (BCJ) amplitudes relations~\cite{Bern:2008qj}. Beyond tree level, the differential representation has proven useful for evaluating scalar one-loop Witten diagrams in both AdS~\cite{Herderschee:2021jbi} and dS~\cite{Gomez:2021ujt}.
Originally motivated by the infinite tension limit of certain string theory expressions for scalar external states~\cite{Eberhardt:2020ewh,Roehrig:2020kck},  the differential representation parallels many features of the momentum-space representation of 
flat-space scattering amplitudes~\cite{Diwakar:2021juk}, see also~\cite{Gomez:2021qfd,Sivaramakrishnan:2021srm}. In particular, conformal/$SO(d,2)$ generators correspond to momentum vectors and the conformal Ward identity (CWI) corresponds to momentum conservation. It would be interesting to understand if this similarity points to deeper structures which may be accessed by reformulating the bulk theory on a noncommutative space.

In this paper, we take steps towards proving that the differential representation exists for general AdS field theories with spinning fields in their spectrum. We will show using a Berends-Giele-type recursion \cite{Berends:1987me} that, at tree level, the nonlocality of the differential representation can be put in one-to-one correspondence  with bulk-bulk propagators while at the same time being expressed entirely in terms of boundary data. The remainder, acted upon by these nonlocal operators, is local and involves integration over a single bulk point for arbitrary multiplicity correlators.\footnote{Related approaches, along the lines of~\cite{Mafra:2015vca,Mizera:2018jbh,Garozzo:2018uzj,Bridges:2019siz,Cheung:2021zvb}, may be useful to construct color-kinematics-satisfying representations for boundary correlators, but we will not follow these directions in this paper.}
The remaining step is to organize this local expression as a collection of conformal generators acting on a single contact diagram. We prove that this is always possible for scalar AdS boundary correlators and illustrate it in nontrivial examples.
We find the differential representation of four-point scalar correlators due to vector and graviton exchange. Comparison with the flat space scattering amplitudes in the relevant theories reveals a close similarity. The vector-mediated correlators can be obtained from flat space amplitudes  by replacing the Mandelstam variables with their corresponding (suitably-ordered) conformal generators and the momentum-conservation constraint with the contact diagram. Graviton-mediated correlators follow a similar, although slightly more subtle, pattern. 
We also outline general properties of the local factor of boundary correlators at arbitrary multiplicity and show that in the limit of large dimension, all scalar boundary correlators can be obtained from flat space amplitudes via this replacement.

While the differential representation is extremely useful for understanding (differential) properties of boundary correlators, the relation between the differential representation and explicit functions, e.g. in position space, is nontrivial because of the presence of nonlocal differential operators. We provide a general strategy for converting the former into a linear combination of contact diagrams; this method relies only on the properties of contact diagrams under the action of the generators of the conformal group.

The paper is organized as follows. 
In section~\ref{revembspace}, we review aspects of the embedding space formalism in the presence of spinning states. 
In section~\ref{reviewdiffrep},  review aspects of the differential representation of boundary correlators and construct it explicitly for scalar theories with nonderivative interactions.
In section~\ref{BGsection}, show how to separate the local and nonlocal part of each diagram contributing to a boundary correlator, that for general spinning states the resulting nonlocal operators are in one-to-one correspondence with bulk-bulk propagators while simultaneously acting solely on boundary points and that the correponding local part is given in terms of standard Feynman vertex factors. The construction described in this section applies more generally to generic AdS Green's functions, with sources placed at both bulk and boundary points.
Assuming that certain differential operators defined in Ref.~\cite{Sivaramakrishnan:2021srm}  exist, the construction in this section, based on the Berends-Giele recursion, applies to general smooth curved spaces.
In section~\ref{sec:scalar}, we give examples of correlators mediated by vector and graviton exchange to illustrate that the remaining local part can indeed be organized in terms of a single contact integral acted 
upon by AdS symmetry generators and the correspondence with flat space scattering amplitudes.
Section~\ref{sec:proof} contains a recursive proof that the local part of scalar boundary correlators has the form required by the differential representation.  
In section~\ref{sec: evaluationofdiffrep}, we provide a general strategy for converting the differential representation to a sum of $D$-functions and section~\ref{sec:conclusion} contains our conclusions and outlook.
Appendix~\ref{sec:graviton} details the AdS graviton propagator in de Donder gauge.

\paragraph{Note added:} During the completion of this project we became aware of concurrent work by Cheung, Parra-Martinez and Sivaramakrishnan~\cite{Cheung:2022pdk} containing some overlap with our work for purely scalar theories.
Based on the evaluation of field one-point function and a judicious reorganization of the equations of motion, differential representations --- referred to as ``isometric representations'' therein --- are constructed for various bi-adjoint scalar theories and the nonlinear sigma model on symmetric spaces. 
%
%
We thank the authors of ref.~\cite{Cheung:2022pdk} for sharing their work with us before submission.

\section{Embedding Space for Spinning States\label{revembspace}}

In this section, we review~\cite{Costa:2014kfa} some useful tools for manipulating symmetric traceless tensors in the embedding space. The $\ads_{d+1}$ spacetime can be realized as a hyperboloid in the embedding space $\mathbb{R}^{d,2}$, defined by the constraint $X^{2}=-R_{\ads}^2=-1$,
\begin{align}
    -(X^0)^2+(X^1)^2+\ldots+(X^d)^2-(X^{d+1})^2=-1\,.
\end{align}
Note that $X^A$ is also the normal vector of this hypersurface.\footnote{Unless otherwise stated, we keep $R_{\ads}=1$ in our formulae.} Points in the bulk that are covered by the Poincar\'{e} patch can be parameterized as
\begin{align}
    X^A\equiv(X^{\mathsf{a}},X^d,X^{d+1})=\frac{1}{z}\Big(x^{\mathsf{a}},\frac{1-x^2-z^2}{2},\frac{1+x^2+z^2}{2}\Big)\,.
\end{align}
Points on the conformal boundary of $\ads_{d+1}$ are identified with the projective null vectors of the embedding space,
\begin{align}
    P^A\equiv\Big(x^{\mathsf{a}},\frac{1-x^2}{2},\frac{1+x^2}{2}\Big)\,.
\end{align}
Lorentz transformations in the embedding space $\mathbb{R}^{d,2}$ can be identified with bulk isometry and boundary conformal transformations. 

We consider a tensor $H_{A_{1}\ldots A_{s}}$ of the embedding space that uniquely defines a symmetric traceless tensor in $\ads_{d+1}$. In addition to being symmetric and traceless itself, $H_{A_1\ldots A_s}$ needs to be transverse to the AdS hypersurface, 
\begin{equation}
\label{transverse}
X^{A_{1}}H_{A_{1}\ldots A_s}(X)=0   \ .
\end{equation}
Instead of working with $H_{A_{1}\ldots A_s}$ directly, we contract its indices with an auxiliary bulk polarization vector $W^{A}$,
\begin{equation}
H(X,W)\equiv W^{A_{1}}\ldots W^{A_{s}}H_{A_{1}\ldots A_{s}}(X)    
\label{Wcontraction}
\end{equation}
which satisfies $W^{2}=W\cdot X=0$. We can uniquely recover the tensor $H_{A_{1}\ldots A_{s}}$ from $H(X,W)$ by acting with the operator~\cite{Costa:2014kfa},
\begin{align}
\label{Koperator}
    K_A&=\frac{d-1}{2}\left[\frac{\partial}{\partial W^A}+X_A\left(X\cdot\frac{\partial}{\partial W}\right)\right]+\left(W\cdot\frac{\partial}{\partial W}\right)\frac{\partial}{\partial W^A}\\
    &\quad +X_A\left(W\cdot\frac{\partial}{\partial W}\right)\!\left(X\cdot\frac{\partial}{\partial W}\right)-\frac{1}{2}W_A\left[\frac{\partial^2}{\partial W\cdot\partial W}+\left(X\cdot\frac{\partial}{\partial W}\right)\!\left(X\cdot\frac{\partial}{\partial W}\right)\right], \nonumber
\end{align}
namely,
\begin{align}
    H_{A_1\ldots A_s}(X)=\frac{1}{s!\left(\frac{d-1}{2}\right)_s}K_{A_1}\ldots K_{A_s}H(X,W)\,,
\end{align}
where $(a)_n\equiv\frac{\Gamma(a+n)}{\Gamma(a)}$ is the Pochhammer symbol. The resulting tensor is guaranteed to be symmetric, traceless, and transverse to AdS since $K_A$ satisfies
\begin{align}
     X^AK_A=0\,,\qquad K_A K^A = 0\,, \qquad K_A K_B = K_B K_A\,.
\end{align}
We can also define a polynomial of $K_A$ operators,
\begin{align}
    H(X,K)\equiv\frac{1}{s!\left(\frac{d-1}{2}\right)_s}K^{A_1}\ldots K^{A_s}H_{A_1\ldots A_s}(X)\,,
\end{align}
such that the contraction of two symmetric traceless AdS tensors is given by
\begin{align}\label{eq:contraction}
    H_{A_1\ldots A_s}(X)F^{A_1\ldots A_s}(X)=H(X,K)F(X,W)=F(X,K)H(X,W)\,.
\end{align}
Many equations vastly simplify using the auxiliary polarization vectors $W$. The AdS covariant derivative in this representation is 
\begin{align}
    \nabla_A=G_{A}{}^{B}\frac{\partial}{\partial X^B}-\frac{W_A}{X^2} \left(X\cdot \frac{\partial}{\partial W}\right)\,,
\label{laplacian}    
\end{align}
where $G^{AB}=\eta^{AB}-\frac{X^{A}X^{B}}{X^2}$ is the metric tensor that projects
dynamics onto the AdS hypersurface. The divergence of a tensor can be represented by
\begin{align}\label{eq:divergence}
    (\nabla\cdot H)(X,W)=(\nabla\cdot K)H(X,W)\equiv\frac{1}{s\left(\frac{d-3+2s}{2}\right)}(\nabla_A K^A)H(X,W)\,,
\end{align}
while the Laplacian is simply $(\nabla^2 H)(X,W)\equiv \nabla_A\nabla^A H(X,W)$.

A similar representation exists for spinning boundary correlators. We consider a symmetric traceless tensors $B_{A_{1},\ldots,A_{s}}(P)$ that lives on the boundary of $\ads_{d+1}$. We can similarly contract it with boundary polarization vectors $Z^A$,
\begin{equation}
B(P,Z)=Z^{A_{1}}\ldots Z^{A_{s}}B_{A_{1},\ldots,A_{s}}  \ .  
\end{equation}
where $Z^{2}=Z\cdot P=0$. For spinning boundary states, the physical null polarization $\epsilon^{\mathsf{a}}$ is embedded into $Z$ through
\begin{align}
    Z^A=\Big(\epsilon^\mathsf{a},-\epsilon\cdot x,\epsilon\cdot x\Big)\,.
\end{align}
To uniquely recover $B_{A_1\ldots A_s}$ from $B(P,Z)$, we define the boundary analog of $K_{A}$~\cite{Costa:2011mg}, 
\begin{gather}
\mathscr{D}^{A}=\left (\frac{d}{2}-1+Z\cdot \frac{\partial}{\partial Z} \right ) \frac{\partial}{\partial Z_{A}}-\frac{1}{2}Z^{A}\frac{\partial^{2}}{\partial Z\cdot \partial Z}\,, \nonumber\\
\text{such that}\quad B_{A_{1}\ldots A_{s}}(P)=\frac{1}{s!\left(\frac{d}{2}-1\right)_s}\mathscr{D}^{A_{1}}\ldots \mathscr{D}^{A_{s}}B(P,Z) \ .
\end{gather}

As noted before, the generators of bulk isometry and boundary conformal symmetry are both given by the Lorentz generators of the embedding space,
%
\begin{align}
D_X^{AB}&=\frac{1}{\sqrt{2}}\left[X^{A}\frac{\partial}{\partial X_{B}}-X^{B}\frac{\partial}{\partial X_{A}}+W^{A}\frac{\partial}{\partial W_{B}}-W^{B}\frac{\partial}{\partial W_{A}}\right],\nonumber\\     
D_i^{AB}&=\frac{1}{\sqrt{2}}\left[P_i^{A}\frac{\partial}{\partial P_{i,B}}-P_i^{B}\frac{\partial}{\partial P_{i,A}}+Z_i^{A}\frac{\partial}{\partial Z_{i,B}}-Z_i^{B}\frac{\partial}{\partial Z_{i,A}}\right].     
\label{bulkandbondarygenerators}
\end{align}
An $n$-point boundary correlator $\mathcal{A}(P_i,Z_i)$, or more generally a conformal partial wave, is invariant under conformal transformations. That is, they obey the conformal Ward identity (CWI), 
\begin{equation}
\label{CWI}
\sum_{i=1}^nD_{i}^{AB}\mathcal{A}(P_{i},Z_{i})=0 \,.
\end{equation}
While this name is justified from the perspective of the boundary conformal field theory, we will refer to eq.~\eqref{CWI} as a "conformal Ward identity" also from the perspective of the bulk AdS theory.

The CWI is analogous to the flat space momentum conservation of scattering amplitudes. In particular, the eigenvalue equation for the quadratic Casimir operator $D_i\cdot D_i$,
\begin{align}\label{eq:casimir}
    -D_i\cdot D_i &= P_i\Cdot\frac{\partial}{\partial P_i}\!\left(d{+}P_i\Cdot\frac{\partial}{\partial P_i}\right)+ Z_i\Cdot\frac{\partial}{\partial Z_i}\!\left(d{-}2{+}Z_i\Cdot\frac{\partial}{\partial Z_i}\right)+2Z_i\Cdot\frac{\partial}{\partial P_i}\!\left(P_i\Cdot\frac{\partial}{\partial Z_i}\right)\nonumber\\
    &\cong\Delta_i(\Delta_i-d)+s(s+d-2)\equiv\qc{i,s}\,,
\end{align}
is the analog of ``on-shell mass'' for a boundary state with spin $s$. A scalar has zero spin, so $\qc{\rm scalar}=\Delta(\Delta-d)$. 
The $\cong$ indicates that this identity holds only when the operator acts on a conformal partial wave that has the support of CWI.
For a massless vector (gluon), we have $\Delta=d-1$ and $s=1$, such that the quadratic Casimir vanishes, $\qc{\rm gluon}=0$. On the other hand, for a graviton, we have $\Delta=d$ and $s=2$, such that $\qc{\rm graviton}=2d$.\footnote{The eigenvalue of a quadratic Casimir is \emph{not} the mass parameter that appears in a bulk Lagrangian or in the equation of motion; see e.g.~\cite{Gunaydin:1998km, Gunaydin:1998sw} for a definition of masslessness in AdS.} 
For later convenience we define
\begin{align}
    \DD_{ij}\equiv (D_i+D_j)^{AB}(D_i+D_j)_{AB}\,,\qquad \DD_{I}\equiv \Big(\sum_{i\in I}D_i\Big){}^{AB}\Big(\sum_{i\in I}D_{i}\Big)_{AB}\,.
\end{align}
These operators satisfy the same linear relation as flat space Mandelstam variables on the support of the CWI. For example, at four points, we have
\begin{gather}
\DD_{12}\cong \DD_{34}\,,\quad \DD_{13}\cong \DD_{24}\,,\quad \DD_{23}\cong \DD_{14}\,,\nonumber\\
\label{eq:cwi4}
(\DD_{12}+\DD_{23}+\DD_{13})\cong -\sum_{i=1}^4\qc{i,s_i}\,.
\end{gather}
However, unlike Mandelstam variables, these operators do not necessarily commute with each other. More precisely, $\DD_{I_1}$ and $\DD_{I_2}$ commute if and only if the set $I_1$ and $I_2$ are disjoint or one set completely contains the other.

\section{Differential representation of boundary correlators}\label{reviewdiffrep}

The \emph{differential representation} of an $n$-point boundary correlator takes the form\footnote{As discussed in the introduction, we assume that our boundary correlators have a perturbative Witten diagram expansion in the bulk. From a holographic point of view, they correspond to correlators in a putative boundary CFT in the strong coupling limit.} 
\begin{equation}
\mathcal{A}_{n}=\hat{\mathcal{A}}_{n}D_{\Delta_1,\Delta_2,\ldots,\Delta_n}
\label{eq:diff0}
\end{equation}
where $\hat{\mathcal{A}}_{n}$ is a collection of local and nonlocal differential operators that contains information on boundary states. For example, for 
external spinning states, it contains information on their spin and polarization.  
Regardless of the spin of external states, $\hat{\mathcal{A}}_{n}$ 
acts on a scalar contact diagram ($D$-function), 
\begin{equation}
D_{\Delta_1,\Delta_2,\ldots,\Delta_n}=\begin{tikzpicture}[baseline={([yshift=-.5ex]current bounding box.center)},every node/.style={font=\scriptsize}]
\pgfmathsetmacro{\r}{0.75}
\draw [] (0,0) circle (\r cm);
\filldraw (\r,0) circle (1pt);
\filldraw (60:\r) circle (1pt) node[above right=-2pt]{$\Delta_4$};
\filldraw (120:\r) circle (1pt) node[above left=-2pt]{$\Delta_3$};
\filldraw (-\r,0) circle (1pt) node[left=0pt]{$\Delta_2$};
\filldraw (-120:\r) circle (1pt) node[below left=-2pt]{$\Delta_1$};
\filldraw (-60:\r) circle (1pt) node[below right=-2pt]{$\Delta_n$};
\filldraw (0,0) circle (1pt);
\draw [thick] (-\r,0) -- (\r,0) (60:\r) -- (-120:\r) (120:\r) -- (-60:\r);
\foreach \x in {15,5,-5,-15}
\filldraw (\x:\r+0.2) circle (0.5pt);
\end{tikzpicture}=\int_{\ads}dX \prod_{i=1}^n E_{\Delta_{i}}(X,P_{i})   \ .
\label{contactint}
\end{equation}
Here $E_{\Delta_i}(X,P_i)=\frac{1}{(-2P_i\cdot X)^{\Delta_i}}$ is the scalar bulk-boundary propagator.\footnote{In this work, we omit the overall normalization $\mathcal{C}_{\Delta}$ of a bulk-boundary propagator. Since for a given boundary correlator all bulk-boundary propagators are the same, the omitted factors can only alter the overall normalization.} The $D$-function thus defined provides the support of CWI to the boundary correlator, and plays a role similar to the momentum conservation constraint in the flat space $S$ matrix. 
Generally, up to subtleties related to the noncommutativity of conformal generators, $\hat{\mathcal{A}}_n$ takes a form similar to the corresponding flat space amplitude. With examples and justification to be detailed in later sections, we claim that a generic differential correlator $\hat{\mathcal{A}}_n$ can be written as 
\begin{align}\label{eq:diffA}
    \hat{\mathcal{A}}_n=\sum_{g\in\text{cubic}}C_g\prod_{I\in g}\frac{1}{\DD_{I}+\qc{I,s}}\hat{N}_g\, ,
\end{align}
where $\hat{N}_g$ is a local operator. 
The summation runs over all the trivalent Witten diagrams. Importantly, the bulk-bulk propagator associated with an internal leg $I$ is given by an inverse differential operator,
\begin{align}\label{eq:diffprop}
    I\;\scalebox{1.25}{$\Bigg\{$}\;\begin{tikzpicture}[baseline={([yshift=-0.5ex]current bounding box.center)},every node/.style={font=\scriptsize}]
    \pgfmathsetmacro{\r}{1}
    \pgfmathsetmacro{\v}{0.5}
			\draw [] (0,0) circle (\r cm);
			\tikzset{decoration={snake,amplitude=.4mm,segment length=1.5mm,post length=0mm,pre length=0mm}}
			\filldraw (45:\r) circle (1pt);
			\filldraw (135:\r) circle (1pt);
			\filldraw (-135:\r) circle (1pt);
			\filldraw (-45:\r) circle (1pt);
			\filldraw (165:\r) circle (1pt);
			\filldraw (195:\r) circle (1pt);
			\filldraw (15:\r) circle (1pt);
			\filldraw (-15:\r) circle (1pt);
			\filldraw (-\v,0) circle (1pt) (\v,0) circle (1pt);
			\draw [thick] (-\v,0) -- (\v,0);
			\draw [thick] (45:\r) -- (\v,0);
			\draw [thick] (-45:\r) -- (\v,0);
			\draw [thick] (135:\r) -- (-\v,0);
			\draw [thick] (-135:\r) -- (-\v,0);
			\draw [thick] (165:\r) -- (-\v,0);
			\draw [thick] (195:\r) -- (-\v,0);
			\draw [thick] (15:\r) -- (\v,0);
			\draw [thick] (-15:\r) -- (\v,0);
			\filldraw [draw=black,fill=gray!50!white] (-\v,0) circle (0.2cm);
			\filldraw [draw=black,fill=gray!50!white] (\v,0) circle (0.2cm);
			\node at (0,0) [below=0pt]{$\Delta,s$};
\end{tikzpicture}\;\longrightarrow\frac{1}{\DD_I+\qc{I,s}}\,,
\end{align}
where $\DD_I$ is the quadratic Casimir operator of the AdS symmetry group acting on all external points connected to one end of that internal line and $\qc{I,s}$ is the eigenvalue of the quadratic Casimir for the representation of the internal state. The inverse differential operator $(\DD_I+\qc{I,s})^{-1}$ will be called a propagator in the context of differential representation. We stress that this form is also expected for theories with quartic and higher-point interaction vertices and, if necessary, such vertices are trivially resolved into trivalent ones by multiplying and dividing 
by suitable operators.
For theories with fields charged under a non-abelian gauge or flavor symmetry, $C_g$ in eq.~\eqref{eq:diffprop} is the color factor associated with the diagram. The (local) operator $\hat{N}_g$, acting directly on the $D$-function, can be understood as the analog of the flat-space kinematic numerator factor.

To justify eq.~\eqref{eq:diffA} and in particular the appearance of the nonlocal operators~\eqref{eq:diffprop}, we start with the differential representation of the bulk-bulk propagator, which was first proposed for bi-adjoint $\phi^3$ theory in Ref.~\cite{Eberhardt:2020ewh}.
As the simplest example, we consider the following s-channel scalar Witten diagram
\begin{align}\label{finalintegrandexp}
\begin{tikzpicture}[baseline={([yshift=-.5ex]current bounding box.center)}]
\draw pic at (0,0) {sexchange};
\end{tikzpicture}\!\!=\int_{\ads} dXdY&E_{\Delta_{3}}(X,P_{3})E_{\Delta_{4}}(X,P_{4})\prop_{\Delta}(X,Y)E_{\Delta_{1}}(Y,P_{1})E_{\Delta_{2}}(Y,P_{2})\, ,
\end{align}
and rewrite the bulk-bulk propagator as the inverse of the free-field operator acting on the delta function
\begin{align}
    \mathcal{G}_{\Delta}(X,Y)=-\frac{1}{\nabla_{X}^2-\Delta(\Delta-d)}\delta(X,Y)\,.
\end{align}
We can now integrate out the bulk point $Y$ and get
\begin{align}
    \begin{tikzpicture}[baseline={([yshift=-.5ex]current bounding box.center)}]
\draw pic at (0,0) {sexchange};
\end{tikzpicture}=\int_{\ads}dX&E_{\Delta_3}(X,P_3)E_{\Delta_4}(X,P_4)\nonumber\\[-1.2em]
&\times\frac{(-1)}{\nabla^2-\Delta(\Delta-d)}E_{\Delta_1}(X,P_1)E_{\Delta_2}(X,P_2)\,.
\end{align}
When acting on a product of scalar bulk-boundary propagators, the AdS Laplacian $\nabla^2$ can also be written 
as $\nabla^2=-D_X\cdot D_X$. Then, the identity
\begin{align}
    D_X\cdot D_X \Big[E_{\Delta_1}(X,P_1)E_{\Delta_2}(X,P_2)\Big]=\DD_{12}\Big[E_{\Delta_1}(X,P_1)E_{\Delta_2}(X,P_2)\Big] \ ,
\end{align}
which is a consequence of the conformal Ward identity, converts the quadratic Casimir operator for the bulk point $X$ to the quadratic Casimir operator for the pair $(1,2)$ of boundary points. This gives the differential representation of the s-channel Witten diagram:
\begin{equation}\label{eq:sexchange}
\begin{tikzpicture}[baseline={([yshift=-.5ex]current bounding box.center)}]
\draw pic at (0,0) {sexchange};
\end{tikzpicture}=\frac{1}{\DD_{12}+\Delta(\Delta-d)}D_{\Delta_1,\Delta_2,\Delta_3,\Delta_4} \,.
\end{equation}
It is exactly analogous to the corresponding scattering amplitude in flat space. The derivation for higher point scalar Witten diagrams is similar. We note that the operators $\DD_I$ associated with the internal lines all commute with each other, so there is no ordering ambiguity in the propagators of the differential representation. 

The above derivation applies specifically to scalar intermediate states. In the next section we will generalize eq.~\eqref{eq:sexchange} by showing that the bulk-bulk propagators for a generic spinning state in a generic Witten diagram 
is equivalent to the inverse differential operators in eq.~\eqref{eq:diffprop}. 
More precisely, we will show that each Witten diagram is exactly given by the action of such inverse operators on contact integrals. 
However, we emphasize that the implication of eq.~\eqref{eq:diffA} is stronger: it requires that all the aforementioned contact integrals can be realized as local operators $\hat{N}_g$ acting on a single $D$-function. Although we do not have a general proof of this stronger statement, it is supported by several nontrivial examples that involve derivatively coupled fields, for example NLSM~\cite{Diwakar:2021juk} and Yang-Mills-Chern-Simons on $\ads_{3}$~\cite{Roehrig:2020kck}. In section~\ref{sec:scalar}, we will extend this list to include scalars minimally coupled to gluons and gravitons.

Because of its similarity to flat space amplitudes that goes beyond that of e.g. Mellin space or AdS momentum representations, the differential representation~\eqref{eq:diffA} can potentially manifest certain hidden structures in the boundary correlators, one of which is color-kinematics duality. For this purpose, we restrict ourselves to theories with only AdS-massless states.\footnote{See e.g.~\cite{Gunaydin:1998km, Gunaydin:1998sw} for a general definition of masslessness in AdS space.} 
As discussed in~\cite{Diwakar:2021juk}, with a correlator organized as in eq.~\eqref{eq:diffA}, color-kinematics duality is formally defined as in flat space.
For example, at four points, a color dressed correlator is said to obey the duality if it can be written in the form,
\begin{align}\label{eq:ckdual}
    \mathcal{A}_4=\left[C_{\rm s}\frac{1}{\DD_{12}}\hat{N}_{\rm s}+C_{\rm t}\frac{1}{\DD_{23}}\hat{N}_{\rm t}+C_{\rm u}\frac{1}{\DD_{13}}\hat{N}_{\rm u}\right]D_{d,d,d,d}\,,
\end{align}
where the kinematics numerators satisfy the same Jacobi identity as the color factors,
\begin{align}
    C_{\rm s}+C_{\rm t}+C_{\rm u}=0\,,\qquad \hat{N}_{\rm s}+\hat{N}_{\rm t}+\hat{N}_{\rm u}=0\,.
\end{align}
We emphasize that the kinematic Jacobi is now an operator relation. For scalar boundary states, the kinematic numerators consist of $\DD_{ij}$ operators, while further operators may also be needed for spinning boundary states. 
The existence of the above BCJ representation is nontrivial and is currently a conjecture even for theories that exhibit color-kinematics duality in flat space. However, if a correlator does obey the duality, then eq.~\eqref{eq:diffA} immediately implies nontrivial differential relations among color ordered correlators~\cite{Diwakar:2021juk}. For the four-point example in eq.~\eqref{eq:ckdual} it yields
\begin{equation}\label{eq:bcj}
\DD_{12}A(1,2,3,4)=\DD_{13}A(1,3,2,4) \, .  
\end{equation}
The derivation is the same as that for the BCJ amplitudes relations of flat space amplitudes, and can be easily generalized to generic cases. 
The NLSM correlators obey color-kinematics duality up to at least six points~\cite{Diwakar:2021juk}. It was also shown in Ref.~\cite{Diwakar:2021juk} that the four-gluon color-ordered correlators satisfy the differential BCJ relation~\eqref{eq:bcj}. 
A color-kinematics-satisfying representation of this correlator, as eq.~\eqref{eq:ckdual}, is yet to be found, but remains plausible.
In addition to exposing color-kinematics duality, the differential representation will likely be instrumental in revealing the existence of double copy relations for AdS correlators.\footnote{A particular double copy of NLSM correlators into special Galileon-like theory in $\ads_4$ is discussed in~\cite{Sivaramakrishnan:2021srm}.}

We remark that the straightforward connection between color-kinematics duality and BCJ relations is one of the most attractive features of the differential representation. This is due to the fact that in the massless limit the CWI takes the same form and has the same consequences as flat space momentum conservation for massless kinematics. 
This connection is absent in momentum-space and Mellin-space boundary correlators that respect the color-kinematics duality. For example, in the momentum space, the obstruction to a BCJ relation is a contact term proportional to the total energy, which is caused by the non-conservation of the momentum in the radial direction~\cite{Armstrong:2020woi,Albayrak:2020fyp}. 
In Mellin space, the proliferation of massive poles in Mandelstam variables also obscures the existence of a simple amplitude relation~\cite{Zhou:2021gnu}.
It would be interesting to understand manifestation of eq.~\eqref{eq:bcj} and its higher-point extension~\cite{Diwakar:2021juk} in these formulations of AdS boundary correlators.

\section{Isolating nonlocalities of AdS correlation functions
}
\label{BGsection}

An essential step towards constructing differential representations for AdS correlators is expressing all bulk-bulk propagators as inverse differential operators acting on the external labels and identifying the remainder. To also capture their origin in particle exchange, we will refer to these operators as ``nonlocalities''. We will refer to the remainder as ``the integrand'' or ``the local factor'' and its general expression is that of a rational function perhaps acted upon by local differential operators. 
The Berends-Giele recursion in AdS will provide a means to a generic construction of this separation.

The flat space (off-shell) Berends-Giele recursion \cite{Berends:1987me} is an algorithmic construction of higher-point 
Green's functions in terms of lower-point ones and propagators and vertex factors constructed from the Lagrangian. It can also be interpreted as the perturbative construction of solutions to the classical 
equations of motion in the presence of arbitrary sources or, alternatively, as the construction of the one-point function in the presence of those sources~\cite{Monteiro:2011pc}.
In this latter interpretations it has a straightforward generalization to more general spacetimes. 

We now introduce this formulation in the context of computing AdS correlation functions. For a generic field theory on AdS background, we organize the field content into a vector $\Phi\equiv(\phi_{s_1},\phi_{s_2},\dots)^{\rm T}$, where $s_{1,2,\ldots}$ are the spins of the corresponding fields. The free equation of motion for $\Phi$ can be described by a diagonal matrix of operators $\mathbb{O}\equiv{\rm diag}(O_{s_1}, O_{s_2},\dots)$, and the interaction is given by some Lagrangian~$\mathcal{L}_{\rm int}$. Suppose we couple the system to the sources $J^{(0)}\equiv (J^{(0)}_{s_1}, J^{(0)}_{s_2},\dots)^{\rm T}$; we can formally write the full equation of motion as
\begin{align}
    \mathbb{O}_X \Phi(X) - \frac{\delta {\cal L}_{\rm int}}{\delta \Phi}(X) 
    = -\Pi J^{(0)}(X) \ ,
    \label{eom}
\end{align}
where $\Pi J^{(0)}\equiv(\Pi_{s_1}\circ J_{s_1}^{(0)},\Pi_{s_2}\circ J_{s_2}^{(0)},\ldots)^{\rm T}$ projects the sources to their symmetric and traceless components. Here $\Pi_{s}$ is the spin-$s$ projector, and the circle $\circ$ denotes the implicit index contraction.
We can solve the equation of motion iteratively,
\begin{align}
    \Phi(X)=\sum_{k=1}^{\infty}\Phi^{(k)}(X)\,,
\end{align}
where $\Phi^{(k)}$ couples to $k$ source vectors $J^{(0)}$. Assuming the sources are localized, we can write the formal solution coupling to $k$ sources as
\begin{align}
    \Phi^{(1)}(X, Y_1) &= -\mathbb{O}^{-1}_X \Pi J^{(0)}(X, Y_1)
    \\
    \Phi^{(k)}(X, \bm Y_k) &= \left.\mathbb{O}^{-1}_X \Pi \frac{\delta {\cal L}_{\rm int}}{\delta \Phi}(X)\right|_{\Phi(X)\rightarrow \sum_{i=1}^{k-1}
    \Phi^{(i)}(X, \bm Y_i)} ^{k \text{ factors of } J^{(0)}} 
    \nonumber\\
    &\equiv -\mathbb{O}^{-1}_X \Pi J^{(k)}(X, \bm Y_k) \ ,
    \label{BGrecursion_operator_form}
\end{align}
where $\bm Y_k\equiv\{Y_1,Y_2,\ldots,Y_k\}$ is the set of $k$ source locations. The restriction to ``$k$ factors of $J^{(0)}$'' instructs one to pick the term with $k$ sources after the variation of the interaction Lagrangian is evaluated on the solution with $k-1$ sources, and 
the last line defines the current $\Pi J^{(k)}(X, \bm Y_k)$, which is also the $k$-source term of the one-point function of $\Phi$.
Here $\mathbb{O}^{-1}$ is the inverse of the operator $\mathbb{O}$ and is to be interpreted as the negative of the Green's function of that operator, i.e. the diagonal matrix 
of Green's functions for the operators $O_{s_i}$. 
Thus, restoring the position dependence and denoting this Green's function by 
$\mathbb{G}(X,U)$, the solution is
\begin{align}
    \Phi^{(1)}(X, Y_1) &= \int dU \, \mathbb{G}(X,U) \circ \Pi J^{(0)}(U, Y_1)\,,
    \\
    \Phi^{(k)}(X, \bm Y_k) &= \int dU \, \mathbb{G}(X,U) \circ\Pi
        \left.\frac{\delta {\cal L}_{\rm int}}{\delta \Phi}(U)\right|_{\Phi(U)\rightarrow \sum_{i=1}^{k-1}\Phi^{(k)}(U, \bm Y_i)} ^{k \text{ factors of } J^{(0)}} 
        \nonumber\\ 
        &= \int dU \, \mathbb{G}(X,U) \circ \Pi
        J^{(k)}(U, \bm Y_k) \ ,
         \label{BGrecursion_positionspace_form}
\end{align}
where $\circ$ denotes implicit index contraction between Green's functions and vertices and sources, and $\bm Y_k$ is the vector of location of $k$  sources as before. The correlation function with $k+1$ bulk points is then obtained by trivially differentiating with respect to the $k$ sources.\footnote{In flat space, scattering amplitudes are obtained by replacing the sources by $\mathbb{O}$ 
acting on solutions of the free equations of motion (i.e. the external legs are ``amputated'', as instructed by the LSZ reduction).} 
In AdS space, placing the sources $J^{(0)}$ on the boundary yields 
the standard position-space form or Witten diagrams. It will also bring us closer 
to constructing a differential representation for AdS correlation functions. 

To this end, it will be useful to first discuss the free equation of motion for 
spin-$s$ fields in AdS space and their properties vis-\`a-vis the AdS symmetry group. 
Such fields are described by symmetric traceless $s$-index tensors; for a uniform treatment we will contract their free indices as well as those of sources 
with null vectors $W$, as in eq.~\eqref{Wcontraction}, so all fields and sources 
are formally scalars; we will refer to them as $\phi_s(X,W)$ and $J_s^{(0)}(X, W)$, respectively. If necessary, the 
tensors can be recovered by stripping off the null vectors using a suitable product 
of $K$ operators defined in eq.~\eqref{Koperator}. Index contract is realized as in eq.~\eqref{eq:contraction}. We can also drop the symmetric and traceless projector $\Pi$ since the action of $K$ on $W$ automatically enforces these properties.
We will assume that the gauge symmetries have been fixed such that the free equation of motion is given by the operator ${O}_s = (\nabla_X^2-m_s^2)$, so the Green's function equation is~\cite{Costa:2014kfa}
\begin{gather}
    (\nabla_X^2-m_s^2)G(X,W_X,U,W_U)=-\delta(X,U)(W_X\cdot W_U)^j \,,\\
    m_s^2=\Delta(\Delta-d)-s\,.\nonumber
\end{gather}
Here, the covariant derivative $\nabla_{X,A}$ is given in eq.~\eqref{laplacian},
and $m_s$ is the mass parameter in the action. ``Massless'' spin-$s$ 
fields in AdS~\cite{Fronsdal:1978rb} are given by $\Delta=s+d-2$ such that $m_s^2 =(s+d-2)(s-2)-s$.\footnote{Scalar fields are an exception to this rule, obeying $\Delta=d$.} In general, however, $m_s$ can be arbitrary; if the AdS theory has a string theory interpretation, $m_s^2$ is related to the mass of string states
and, consequently, for massive string states to the dimension of operators developing anomalous dimensions in the dual CFT.

On symmetry grounds, one might expect that the quadratic operator is closely related to the quadratic Casimir of $SO(d, 2)$. We will see that this is indeed the case.
Using the transversality properties \eqref{transverse} of AdS fields in embedding 
space,
\begin{align}
    X\cdot \frac{\partial}{\partial W}\phi_s(X,W)=0\,,\quad W\cdot\frac{\partial}{\partial W}\phi_s(X,W)=s\phi_s(X,W)\,,
\end{align}
one can show that $\nabla^2$ and $-D_X\cdot D_X$, the quadratic Casimir of $SO(d, 2)$ at position $X$,
are indeed very similar,
\begin{align}
    \nabla_X^2 \phi_s(X,W) &=\partial_A(G^{AB}\partial_B)\phi_s(X,W)-s \phi_s(X,W) \ ,\nonumber\\
    -D_X\cdot D_X \phi_s(X,W)& =\partial_A(G^{AB}\partial_B)\phi_s(X,W)+s(s+d-2)\phi_s(X,W) \ ,
\end{align}
where the generators of $SO(d, 2)$ are defined in eq.~\eqref{bulkandbondarygenerators},
and include actions on the spin degrees of freedom. 
Thus, the quadratic operator can be written as
\begin{align}
    (\nabla_X^2-m_s^2)\phi_s(X,W)=\Big(-D_X\cdot D_X-\mathsf{M}_s^2\Big)\phi_s(X,W)
    \label{rewrite}
\end{align}
where $\qc{s} = m_s^2 +s(s+d-1)$. For ``massless'' spin-$s$ 
fields in AdS, this becomes 
$\qc{s} = 2(s-1)(s+d-2)$, i.e. the quadratic 
Casimir for a $s$-index symmetric traceless representation.\footnote{Massless scalar fields are an exception as for them $m_s^2=0$ and $s=0$, so $\qc{s}=0$.} The relation~\eqref{rewrite} holds for generic bulk field, and we will later use it for the current $J^{(k)}$ in eq.~\eqref{BGrecursion_positionspace_form}.

The close relation between the quadratic operator for spin-$s$ 
fields and the quadratic Casimir of the corresponding representation 
together with the conformal Ward identity
are key to organizing the operator formulation of the Berends-Giele recursion in eq.~\eqref{BGrecursion_operator_form} in a form in which all nonlocalities represented 
by inverse operators depend only on the position of the external sources.\footnote{This is the AdS analog of the flat space feature that all propagators in tree-level Feynman diagrams are determined by the external momenta.}
This will reduce the construction of such AdS correlation functions 
to the construction of ``the integrand''.

To see this we note that because all indices of fields are saturated by null vectors $W$
which transform under AdS symmetry group following eq.~\eqref{bulkandbondarygenerators}, they are all scalars and therefore the vector of currents $J^{(k)}(X, \bm Y_k)$ defined in eq.~\eqref{BGrecursion_operator_form} and appended with $W$ dependence 
is also a vector of scalars. It consequently obeys the conformal Ward identity
\begin{align}
    \Big(D_X^{AB} + \sum_{i=1}^k D_{Y_i}^{AB}\Big)J^{(k)}(X, W,\bm Y_k, \bm W_k) = 0 \ .
\end{align}
Here $D_{Y_i}^{AB}$ are $SO(d,2)$ generators acting on $Y_i$ and include 
action on null vectors $W_i$ (the set of which is denoted by $\bm W_k$) 
saturating the tensor indices at that (possibly boundary) position. It is then 
trivial to see that, as long as $X\ne Y_i$ for all 
$i=1, \dots, k$,
\begin{align}
    D_X\cdot D_X  J^{(k)}(X, W,\bm Y_k, \bm W_k) 
    = 
    \DD_{Y_1\ldots Y_k}  J^{(k)}(X, W, \bm Y_k, \bm W_k) \ .
    \label{eq:DxToDijk}
\end{align} 
From here, eq.~\eqref{rewrite} and the definition of $\mathbb{O}_X$, it follows that 
\begin{align}
   & \quad - \mathbb{O}_X^{-1}\ J^{(k)}(X, W,\bm Y_k, \bm W_k)\nonumber\\
    &=
    {\rm diag}\left(\frac{1}{\DD_{Y_1\dots Y_k}+\mathsf{M}_{s_1}^2}, \frac{1}{\DD_{Y_1\dots Y_k}+\mathsf{M}_{s_2}^2}, \dots \right)
     J^{(k)}(X, W, \bm Y_k, \bm W_k)\nonumber\\
     &\equiv \mathbb{O}_{\bm Y_k}^{-1} J^{(k)}(X, W,\bm Y_k, \bm W_k) \ .
\end{align}
Thus, each step in the recursion relation \eqref{BGrecursion_operator_form} involves only quantities that are local with respect to the bulk point $X$,
\begin{align}
    \Phi^{(1)}(X, W, Y_1,W_1) &= \int dU \, \mathbb{G}(X,W,U,K_U)J^{(0)}(U,W_U, Y_1,W_1)\,,
    \\
    \Phi^{(k)}(X,W, \bm Y_k, \bm W_k) &= \mathbb{O}^{-1}_{\bm Y_k} \left.\frac{\delta {\cal L}_{\rm int}}{\delta \Phi}(X,W)\right|_{\Phi(X)\rightarrow \sum_{i=1}^{k-1}
    \Phi^{(i)}(X, \bm Y_i)} ^{k \text{ factors of } J^{(0)}} \nonumber\\
    &=\mathbb{O}^{-1}_{\bm{Y}_k}J^{(k)}(X,W,\bm{Y}_k,\bm{W}_k)\,.
    \label{BGrecursion_better}
\end{align}
In the first step of the recursion, we do not 
introduce $\mathbb{O}^{-1}_{X}$ but write out $\mathbb{G}$ explicitly because they will form the bulk integrand after factoring out all the inverse differential operators related to particle-exchange nonlocalities.
If the source $J^{(0)}$ is inserted at a boundary point $P_1$, these one-point Green's functions become bulk-boundary propagators $\mathbb{E}\equiv\textrm{diag}(E^{s_1},E^{s_2},\ldots)$~\cite{Freedman:1998tz,Mueck:1998wkz},
\begin{align}\label{eq:bdrysource}
    \Phi^{(1)}(X,W,P_1,Z_1)=\mathbb{E}(X,W,P_1,Z_1)j(P_1)\,,
\end{align}
and $j=(j_{s_1},j_{s_2},\ldots)^{\rm T}$ is a vector of distribution functions for boundary sources. The bulk-boundary propagators are part of the definition of the contact integral in the differential representation, see section~\ref{reviewdiffrep}.

Our Berends-Giele recursion works for both bulk and boundary correlation functions. In the following, we restrict ourselves to boundary correlators, and consider only sources of the form~\eqref{eq:bdrysource}. The $n$-point boundary correlator can then be extracted from $\Phi^{(n-1)}$. We first remove the overall nonlocality $\mathbb{O}_{\bm{Y}_{n-1}}^{-1}=\mathbb{O}_{Y_1\ldots Y_{n-1}}^{-1}$ from $\Phi^{(n-1)}$ and then contract it with the bulk-boundary propagator originated from the last boundary source at $P_n$,
\begin{align}\label{eq:Wj}
    \mathcal{W}[j]&\equiv\int dX j(P_n)\mathbb{E}(X,K,P_n,Z_n)\mathbb{O}_{Y_1\ldots Y_{n-1}}\Phi^{(n-1)}(X,W,\bm{P}_{n-1},\bm{Z}_{n-1})\nonumber\\
    &=\int dX j(P_n)\mathbb{E}(X,K,P_n,Z_n)J^{(n-1)}(X,W,\bm{P}_{n-1},\bm{Z}_{n-1})\,.
\end{align}
What we get is nothing but the generating functional for connected Witten diagrams. Thus, the boundary correlators can be obtained by taking the functional derivatives with respect to the source distribution function $j(P_i)$,
\begin{align}\label{eq:bdrycorr}
    \langle\Phi(P_1,Z_1)\Phi(P_2,Z_2)\ldots\Phi(P_n,Z_n)\rangle=\frac{\delta^{n}\mathcal{W}[j]}{\delta j(P_1)\delta j(P_2)\ldots\delta j(P_n)}\,.
\end{align}
Remarkably, all the $\mathbb{O}^{-1}_{\bm{Y}_k}$ operators contained in $J^{(n-1)}$ act only on $P_n$ or $Z_n$. They can be trivially pulled out of the bulk integral in eq.~\eqref{eq:Wj} such that we are left with integrating out an effective $n$-point contact interaction in the bulk for each Witten diagram. Therefore, we have shown that we can ``factor out'' the nonlocalities due to particle exchange as inverse differential operators that only act on boundary states, while the remaining bulk integral involves only contact interactions.  

To illustrate the recursion, we consider a single spin-$s$ field $\phi$ whose free equation of motion is given by the quadratic operator $O=\nabla^2-m_s^2$. We then introduce both cubic and quartic vertices, ${\cal V}_3(\nabla,G, W_1, W_2, W_3)$ and ${\cal V}_4(\nabla,G, W_1, W_2, W_3, W_4)$. The vertices explicitly depend on the covariant derivative $\nabla_A$ and AdS projector $G^{AB}$. Here, we contract the tensor indices with polarization vectors. For $\mathcal{V}_3$, we contract the two incoming legs with $W_2$ and $W_3$ respectively, and the outgoing leg with $W_1$, and similarly for $\mathcal{V}_4$. We have used the freedom of integration-by-parts such that the covariant derivatives in $\mathcal{V}_3$ and $\mathcal{V}_4$ only act on the incoming currents. Then, 
with the shorthand notation that suppresses the dependence on source polarizations, $\phi^{(k)}_{1\dots k}(X,W)\equiv \phi^{(k)}(X,W, \bm Y_k, \bm W_k)$ and $ \mathbb{O}^{-1}_{ij\dots}\equiv\mathbb{O}^{-1}_{Y_iY_j\dots}$, the first three steps in the recursion yield
\begingroup
\allowdisplaybreaks
\begin{subequations}
\begin{align}
\phi^{(1)}_1(X,W) &= E^{s}(X,W,P_1,Z_1)j(P_1)\,,
\\
\phi^{(2)}_{12}(X,W) &= \mathbb{O}^{-1}_{12} 
{\cal V}_3(\nabla_X, G, W, K_1, K_2) \phi^{(1)}_1(X,W_1)\phi^{(1)}_2(X,W_2)\,,
\\
\phi^{(3)}_{123}(X,W
) &= 
\mathbb{O}^{-1}_{123}
\Big[
{\cal V}_4(\nabla_X,G, W, K_1, K_2, K_3)
\\*
&\quad+
\mathbb{O}^{-1}_{12}
{\cal V}_3(\nabla_X, G, W, K_3, K_x)
{\cal V}_3(\nabla_X, G, W_x, K_1, K_2)
\nonumber\\*
&\quad+
\mathbb{O}^{-1}_{13}
{\cal V}_3(\nabla_X, G, W, K_2, K_x)
{\cal V}_3(\nabla_X, G, W_x, K_1, K_3)
\nonumber\\*
&\quad+
\mathbb{O}^{-1}_{23}
{\cal V}_3(\nabla_X, G, W, K_1, K_x)
{\cal V}_3(\nabla_X, G, W_x, K_2, K_3)
\Big]\prod_{i=1}^{3}\phi^{(1)}_i(X,W_i)\,.
 \nonumber
\end{align}
\end{subequations}
\endgroup
The contraction of $K$ with $\nabla$ in $\mathcal{V}_3$ and $\mathcal{V}_4$ follows eq.~\eqref{eq:divergence}. According to eq.~\eqref{eq:Wj} and~\eqref{eq:bdrycorr}, the three- and four-point boundary correlators are
\begingroup
\allowdisplaybreaks
\begin{align}
\langle\phi_1\phi_2\phi_3\rangle &= 
\int dX {\cal V}_3(\nabla_X, G, K_1, K_2, K_3) \prod_{i=1}^{3}E^s(X,W_i,P_i,Z_i)\,,
\\
\langle\phi_1\phi_2\phi_3\phi_4\rangle &= \int dX 
{\cal V}_4(\nabla_X, G, K_1, K_2, K_3, K_4)\prod_{i=1}^{4}E^s(X,W_i,P_i,Z_i)
\\*
&\hspace{-4em}+
\mathbb{O}^{-1}_{23}\int dX 
{\cal V}_3(\nabla_X, G, K_1, K_4, K_x)
{\cal V}_3(\nabla_X, G, W_x, K_2, K_3)\prod_{i=1}^{4}E^s(X,W_i,P_i,Z_i)
\nonumber\\*
&\hspace{-4em}+
\mathbb{O}^{-1}_{24}\int dX
{\cal V}_3(\nabla_X, G, K_1, K_3, K_x)
{\cal V}_3(\nabla_X, G, W_x, K_2, K_4)\prod_{i=1}^{4}E^s(X,W_i,P_i,Z_i)
\nonumber\\*
&\hspace{-4em}+
\mathbb{O}^{-1}_{34} \int dX
{\cal V}_3(\nabla_X, G, K_2, K_2, K_x)
{\cal V}_3(\nabla_X, G, W_x, K_3, K_4)\prod_{i=1}^{4}E^s(X,W_i,P_i,Z_i)\,.
 \nonumber
\end{align}
\endgroup
All operators $\mathbb{O}^{-1}_{ij}$ are independent of the integration variable, 
so in each term they can be extracted from under the integral. As in flat space, 
each term corresponds to a tree-level graph and the denominators of flat-space propagators translate as
\begin{align}
    s_{i_1\dots i_k} +m_s^2 \mapsto \DD_{i_1\dots i_k} + \qc{s} \ ,
    \label{propreplacement}
\end{align}
with $\qc{s}$ emerging from eq.~\eqref{rewrite}.

We note that for scalar theories with any number of scalars with mass parameters $m_i$ and polynomial non-derivative interactions, the Berends-Giele recursion detailed here implies that the integrands of the corresponding boundary correlators are obtained from the flat space scattering amplitudes via the 
further replacement
\begin{align}
    \delta\Big(\sum_{i=1}^{n} p_i\Big)\mapsto D_{\Delta_1,\dots \Delta_n} \ ,
    \label{deltareplacement}
\end{align}
where $\Delta_i$ is the dimension of the boundary operator dual to the $i$-th 
external field with mass parameter $m_i$, i.e. $m_i^2 = \Delta_i(\Delta_i-d)$.
For single-field theories, we recover the differential representation of boundary correlators in the form derived in section~\ref{reviewdiffrep}.
Moreover, as expected, the tree-level AdS boundary correlators of the cubic biadjoint scalar theory obey color-kinematics duality manifestly, and they also obey correlator relations for any multiplicity, as they do in flat space.

Other theories, such as theories with derivative interactions, are currently 
analyzed on a case by case basis. While the recursion discussed here guarantees 
that boundary correlators can be written as a specific set of inverse differential operators acting on a local expression, it does not immediately guarantee that
local expression can itself be written as a collection of local operators acting 
in a single contact integral $D_{\Delta_1,\dots \Delta_n}$. 
We will however see in later sections that this expectation pans out for
scalars coupled with gauge fields or gravitons. In the examples we will discuss, the replacements \eqref{propreplacement} and \eqref{deltareplacement} will continue to relate flat space amplitudes and boundary correlators. In section~\ref{largeD} and section~\ref{sec:proof} we will prove that all scalar AdS boundary correlators have a differential representation of form given in eqs.~\eqref{eq:diff0} to~\eqref{eq:diffA} with a manifestly local numerators $\hat N_g$ and also see superficial limitations of replacements like $p_i\cdot p_j\rightarrow D_i\cdot D_j$ for relation between the local factor of scalar correlators and flat space scattering amplitudes.  

\section{Scalar correlators mediated by spinning states}\label{sec:scalar}

In this section, we present a remarkably simple differential representation of the four-point correlators of scalars minimally coupled to gluons or gravitons. We will include here explicit examples of the general discussion in the previous section on the separation of the nonlocal and local factors of the correlator, and the organization of the latter in terms of conformal generators acting on a single contact integral.

The gauge theory we consider is given by the Lagrangian
\begin{align}\label{eq:Lgauge}
    \mathcal{L}_{\rm gauge}=-\frac{1}{4}F_{\mu\nu}^aF^{a,\mu\nu}-\frac{1}{2}D_{\mu}\phi^aD^{\mu}\phi^a-\frac{1}{2}D_{\mu}\tilde\phi^aD^{\mu}\tilde\phi^a-\frac{1}{4}f^{abc}f^{cde}\phi^a\tilde\phi^b\phi^d\tilde\phi^e\,,
\end{align}
and the gravitational theory is its double copy.
Here $\phi^a$ and $\tilde\phi^a$ are two distinct massless scalars, and $D_{\mu}\phi^a=\partial_{\mu}\phi^a+f^{abc}A_{\mu}^b\phi^c$. The 
generalization to more than two flavors is straightforward and the 
relevant Lagrangian is the scalar sector of the dimensional reduction 
of YM theory.

\subsection{Flat space amplitudes}

The flat space amplitudes following from the Lagrangian \eqref{eq:Lgauge} enjoy color-kinematics duality,
\begin{subequations}
\begin{align}
    \mathcal{A}_4^{\rm flat}(\phi_1\phi_2\phi_3\phi_4)&=C_{\rm s}\frac{s_{23}-s_{13}}{s_{12}}+C_{\rm t}\frac{s_{13}-s_{12}}{s_{23}}+C_{\rm u}\frac{s_{12}-s_{23}}{s_{13}}\,,\\
    \mathcal{A}_4^{\rm flat}(\phi_1\phi_2\tilde\phi_3\tilde\phi_4)&=C_{\rm s}\frac{s_{23}-s_{13}}{s_{12}}+C_{\rm t}\frac{-s_{23}}{s_{23}}+C_{\rm u}\frac{s_{13}}{s_{13}}\,,
\end{align}
\end{subequations}
where the numerators satisfy $N_{\rm s}+N_{\rm t}+N_{\rm u}=0$ for both cases. Double copy gives the following gravitational amplitudes,
\begin{subequations}
\begin{align}\label{eq:flatdc}
    \mathcal{M}_4^{\rm flat}(\varphi_1\varphi_2\varphi_3\varphi_4)&=\frac{1}{16}\left[\frac{(s_{23}-s_{13})^2}{s_{12}}+\frac{(s_{13}-s_{12})^2}{s_{23}}+\frac{(s_{12}-s_{23})^2}{s_{13}}\right]\,,\\
    \label{eq:scalardc2}
    \mathcal{M}_4^{\rm flat}(\varphi_1\varphi_2\tilde\varphi_3\tilde\varphi_4)&=\frac{1}{16}\left[\frac{(s_{23}-s_{13})^2}{s_{12}}+s_{23}+s_{13}\right]=\frac{1}{8}\frac{s_{23}^2+s_{13}^2-s_{12}^2}{s_{12}}\,.
\end{align}
\end{subequations}
They correspond to the gravitational Lagrangian
\begin{align}\label{eq:gravity2}
    \mathcal{L}_{\rm gravity}=-\frac{\sqrt{-g}}{2}\Big(g^{\mu\nu}\partial_{\mu}\varphi\partial_{\nu}\varphi+g^{\mu\nu}\partial_{\mu}\tilde\varphi\partial_{\nu}\tilde\varphi\Big)\,.
\end{align}
Note that a four-point contact interaction is not generated by the double copy. 

\subsection{AdS correlators}

We now put both actions~\eqref{eq:Lgauge} and~\eqref{eq:gravity2} in the AdS background and compute the scalar boundary correlators. We start with conventional Witten diagrams and systematically arrange the results in terms of products of conformal generators acting on a contact diagram. 
While doing so, we will revisit in more detail aspects discussed in section~\ref{BGsection}. 
As we will see, the final expressions closely related to 
the flat space amplitudes through replacements similar to those in eqs.~\eqref{propreplacement} and \eqref{deltareplacement}. 

\subsubsection{Gluon-mediated correlators}

We start with the AdS boundary correlator of four identical scalars, minimally coupled with gluons. 
The contribution $A_s$ from the s-channel gluon exchange can be written as 
\begin{align}\label{eq:gluon_exchange}
    \gexchange{-0.5ex}{1}{2}{3}{4}=A_s=\int_{\ads}dX_1dX_2 J_A(P_1,P_2,X_1)\prop^{AB}_g(X_1,X_2)J_B(P_3,P_4,X_2)\,,
\end{align}
where $\prop_g^{AB}(X_1,X_2)$ is the gluon bulk-bulk propagator, and the vertex function $J_A^{(ij)}$ is derived from the interaction Lagrangian $\mathcal{L}_{\rm gauge}$ in eq.~\eqref{eq:Lgauge},
\begin{align}
    J_A(P_i,P_j,X)=E_d(X,P_j)\nabla_A E_d(X,P_i)-E_d(X,P_i)\nabla_A E_d(X,P_j)\,.
\end{align}
Following eq.~\eqref{eq:contraction}, we can evaluate the index contraction by acting the operator $K_A$ onto the polarization $W_A$, such that
\begin{align}
    \gexchange{-0.5ex}{1}{2}{3}{4}=\int_{\ads}dX_1dX_2 J(P_1,P_2,X_1,K_1)J(P_3,P_4,X_2,K_2)\prop_{g}(X_1,X_2,W_1,W_2)\,.
\end{align}
In this form, the gluon bulk-bulk propagator satisfies the differential equation~\cite{Costa:2014kfa}
\begin{align}
    (\nabla_1^2+d)\prop_g(X_1,X_2,W_1,W_2)=-\delta(X_1,X_2)(W_1\cdot W_2)\,,
\end{align}
which leads to the identity
\begin{align}
    (\nabla_1^2+d)\int_{\ads}dX_2 J(P_3,P_4,X_2,K_2)\prop_{g}(X_1,X_2,W_1,W_2)=-J(P_3,P_4,X_1,W_1)\,.
\end{align}
The correlator $A_s$ can thus be formally written as
\begin{align}
    \gexchange{-0.5ex}{1}{2}{3}{4}=-\int_{\ads} dX_1 J(P_1,P_2,X_1,K_1)\frac{1}{\nabla_1^2+d}J(P_3,P_4,X_1,W_1)\,.
\end{align}
Notice that the action of $\nabla^2+d$ on $J(P_3,P_4,X,W)$ is equivalent to that of $\DD_{34}$, cf. eqs.~\eqref{rewrite} and \eqref{eq:DxToDijk}:
\begin{align}
    -(\nabla_1^2+d)J(P_3,P_4,X_1,W_1)=\DD_{34}J(P_3,P_4,X_1,W_1)\,.
\end{align}
We can then pull the operator ${\DD^{-1}_{34}}$ out from the bulk integral, and we are left with only simple contact interaction in the bulk, as indicated by the recursion \eqref{BGrecursion_better}. After performing the integral and simplifying the result using relations between $\DD_{ij}$, we arrive at
\begin{align}\label{eq:sgauge}
    \gexchange{-0.5ex}{1}{2}{3}{4}&=\frac{1}{\DD_{34}}\int_{\ads}dX J_A (P_1,P_2,X)G^{AB}J_B(P_3,P_4,X) \nonumber\\
    &=\frac{1}{\DD_{12}}(\DD_{23}-\DD_{13})D_{d,d,d,d}\, ,
\end{align}
with $D_{d,d,d,d}$ defined by \eqref{contactint} with $\Delta_i = d$. The full color-dressed correlator has therefore the differential representation
\begin{align}
    \mathcal{A}(\phi_1\phi_2\phi_3\phi_4)&=\;\gexchange{-0.5ex}{1}{2}{3}{4}\;+\;\gexchange{-0.5ex}{1}{4}{2}{3}\;+\;\gexchange{-0.5ex}{1}{3}{4}{2}\nonumber\\
    &=\left(C_{\rm s}\frac{1}{\DD_{12}}\hat{N}_{\rm s}^{\phi\phi\phi\phi}+C_{\rm t}\frac{1}{\DD_{23}}\hat{N}_{\rm t}^{\phi\phi\phi\phi}+C_{\rm u}\frac{1}{\DD_{13}}\hat{N}_{\rm u}^{\phi\phi\phi\phi}\right)D_{d,d,d,d}\,,
\end{align}
where the kinematic numerators are
\begin{align}\label{eq:sYM}
    \hat{N}_{\rm s}^{\phi\phi\phi\phi}=\DD_{23}-\DD_{13}\,,\qquad \hat{N}_{\rm t}^{\phi\phi\phi\phi}=\DD_{13}-\DD_{12}\,,\qquad \hat{N}_{\rm u}^{\phi\phi\phi\phi}=\DD_{12}-\DD_{23}\,.
\end{align}
They satisfy the operator kinematic Jacobi identity, just like the flat space amplitude. We note that the same conclusion also applies to $\mathcal{A}(\phi_1\phi_2\tilde\phi_3\tilde\phi_4)$. After considering the contact interaction in eq.~\eqref{eq:Lgauge}, we have
\begin{align}
    \hat{N}_{\rm s}^{\phi\phi\tilde\phi\tilde\phi}=\DD_{23}-\DD_{13}\,,\qquad\hat{N}_{\rm t}^{\phi\phi\tilde\phi\tilde\phi}=-\DD_{23}\,,\qquad\hat{N}_{\rm u}^{\phi\phi\tilde\phi\tilde\phi}=\DD_{13}\,.
\end{align}
Consequently, the color-ordered correlators satisfy the differential BCJ relation
\begin{align}
    \DD_{12}A(1,2,3,4)=\DD_{13}A(1,3,2,4)\,,
\end{align}
where $A(1,2,3,4)=A_{\rm s}-A_{\rm t}$ and $A(1,3,2,4)=A_{\rm t}-A_{\rm u}$.

To further illustrate the similarity to flat space amplitudes, we compute the six-point identical-scalar correlator $\mathcal{A}(\phi_1\phi_2\phi_3\phi_4\phi_5\phi_6)$ following the construction given in section~\ref{BGsection}. The relevant Witten diagrams are of three distinct topologies. While one can simply read off the inverse differential operators for bulk-bulk propagators, the kinematic numerators are:
\begingroup
\allowdisplaybreaks
\begin{align}
    \hat{N}\!\left[\begin{tikzpicture}[baseline={([yshift=-.5ex]current bounding box.center)}]
    \draw pic at (0,0) {sf};
    \end{tikzpicture}\right]&=\frac{1}{2}\Big(\DD_{25}\DD_{13}-\DD_{15}\DD_{23}+\DD_{15}\DD_{24}-\DD_{25}\DD_{14}\nonumber\\*[-2em]
    &\quad\quad\quad+\DD_{26}\DD_{14}-\DD_{16}\DD_{24}+\DD_{45}\DD_{23}-\DD_{35}\DD_{24}\Big)\nonumber\\*
    &\quad+(12\rightarrow 34\rightarrow 56\rightarrow 12)+(12\rightarrow 56\rightarrow 34\rightarrow 12)\,,\label{eq:sf6} \\
    \hat{N}\!\left[\begin{tikzpicture}[baseline={([yshift=-.5ex]current bounding box.center)}]
    \draw pic at (0,0) {hl};
    \end{tikzpicture}\right]&=\frac{1}{2}\Big(\!-\DD_{13}+\DD_{23}+\DD_{14}-\DD_{24}+\DD_{15}-\DD_{25}+\DD_{16}-\DD_{26}\Big)\nonumber\\*[-2em]
    &\quad\quad\times\Big(\DD_{45}-\DD_{46}\Big)\,, \label{eq:hl6}  \\
    \hat{N}\!\left[\ctd{-0.5}{1}{2}{3}{4}{5}{6}\right]&=\frac{1}{2}\Big(\DD_{23}-\DD_{13}+\DD_{14}-\DD_{24}\Big)\,.\label{eq:contact6}
\end{align}
\endgroup
In addition, the color factors for these diagrams are
\begin{align}
    C_{\snowflake}&=f^{b_1b_2b_3}f^{b_1a_1a_2}f^{b_2a_3a_4}f^{b_3a_5a_6}\,, \nonumber\\
    C_{\halfladder}&=f^{a_1a_2b_2}f^{b_2a_3b_3}f^{b_3a_4b_4}f^{b_4a_5a_6}\,, \\
    C_{\cont}&=f^{a_1a_2b_1}f^{a_3a_4b_2}(f^{a_5b_1c}f^{cb_2a_6}+f^{a_5b_2c}f^{cb_1a_6})\,.\nonumber
\end{align}
All other Witten diagrams can be obtained by permutations of the external legs. 
The results are again identical to the flat space amplitudes, up to the replacements $s_{ij}\rightarrow \DD_{ij}$.

If we absorb the contact numerators~\eqref{eq:contact6} into the half-ladder ones~\eqref{eq:hl6}, the numerators satisfy color-kinematics duality. Here we give the numerators for half-ladder diagrams,
\begin{align}
    \hat{N}_{\text{BCJ}}\!\left[\begin{tikzpicture}[baseline={([yshift=-.5ex]current bounding box.center)}]
    \draw pic at (0,0) {ddmhl};
    \end{tikzpicture}\right]=\hat{N}\!\left[\begin{tikzpicture}[baseline={([yshift=-.5ex]current bounding box.center)}]
    \draw pic at (0,0) {hl};
    \end{tikzpicture}\right]+\DD_{123}\,\hat{N}\!\left[\ctd{-0.5}{5}{6}{1}{2}{3}{4}\right] \ ;
\end{align} 
the numerators of the other diagrams can be obtained by using Jacobi identities. In this BCJ representation, the snowflake numerators will differ from eq.~\eqref{eq:sf6} by some additional commutators $[\DD_{ij},\DD_{jk}]$. These commutators are nonzero as operators but annihilate the $D$-functions. Thus, they do not change the boundary correlator. This phenomenon has already been observed for NLSM~\cite{Diwakar:2021juk}. Finally, we note that the differential form of the six-point fundamental BCJ relation
\begin{align}
    0&=\DD_{12}A(1,2,3,4,5,6)+(\DD_{12}+\DD_{23})A(1,3,2,4,5,6)\nonumber\\
    &\quad -(\DD_{25}+\DD_{26})A(1,3,4,2,5,6)-\DD_{26}A(1,3,4,5,2,6) \ ,
\end{align}
holds as a consequence of the color-kinematics duality.

\subsubsection{Graviton-mediated four-point correlator \label{gravexchange}}

We proceed to discuss the four-point AdS boundary correlator due to graviton exchange following from the action~\eqref{eq:gravity2}. While we could directly apply the results of section~\ref{BGsection}, we will proceed as in the case of gluon-mediated four-point AdS boundary correlator and describe the various intermediate steps of the calculation. 

The s-channel graviton exchange is given by~\cite{Costa:2014kfa},
\begin{align}\label{eq:sgraviton}
    \hexchange{-0.5ex}{1}{2}{3}{4}\,
    &=\frac{1}{4}\int_{\ads}dX_1dX_2 T_{AB}(P_1,P_2,X_1) \prop_h^{AB,CD}(X_1,X_2) T_{CD}(P_3,P_4,X_2)\,,
\end{align}
where the vertex function is the scalar stress tensor
\begin{align}
    T_{AB}(P_i,P_j,X)&=\nabla_A E_d(X,P_i)\nabla_B E_d(X,P_j)+\nabla_B E_d(X,P_i)\nabla_A E_d(X,P_j)\nonumber\\
    &\quad -G_{AB}\nabla_CE_d(X,P_i)\nabla^C E_d(X,P_j)\,,
\end{align}
and $\prop_h^{AB,CD}(X_1,X_2)$ is the graviton bulk-bulk propagator. As for gluon exchange, we formally replace $\prop_h^{AB,CD}$ by the inverse of the free equation of motion acting on a delta function. When acting on a vertex function, $\nabla^2$ has a linear relation with the operator $\DD_{ij}$ associated with the external particles sourcing the vertex. Using this relation to trade the former for the latter, we are left with a contact interaction in the bulk, which can be integrated straightforwardly. 

Of course, compared with gluons, gravitons in the bulk are more subtle: they contain both a spin-2 and spin-0 component. We relegate the detailed derivation to appendix~\ref{sec:graviton} and give here the remarkably simple result,
\begin{align}\label{eq:bulkbulks}
    \hexchange{-0.5ex}{1}{2}{3}{4}\,=\frac{1}{4}\frac{1}{\DD_{12}+2d}\int_{\ads}dX T_{AB}(P_1,P_2,X)\Pi^{AB,CD}_{h}(X)T_{CD}(P_3,P_4,X)\,,
\end{align}
where 
\begin{align}
    \Pi^{AB,CD}_{h}(X)=\frac{1}{2}\left(G^{AC}G^{BD}+G^{AD}G^{BC}-\frac{2}{d-1}G^{AB}G^{CD}\right)\,.
\end{align}
We then carry out the bulk integral and write the answer in terms of the differential operators acting on the contact diagram,
\begin{align}\label{eq:bulkint}
    &\int_{\ads}dX T_{AB}(P_1,P_2,X)\Pi^{AB,CD}_{h}(X)T_{CD}(P_3,P_4,X)\nonumber\\
    &=\frac{1}{2}\left(\DD_{14}\DD_{23}+\DD_{13}\DD_{24}-\DD_{12}\DD_{34}\right)D_{d,d,d,d}=\frac{1}{2}(\DD_{13}^2+\DD_{23}^2-\DD_{12}^2)D_{d,d,d,d}\,.
\end{align}
Again, this result displays a very clear analogy with the corresponding flat space amplitude --- the numerators are related by the replacement $s_{ij}\rightarrow \DD_{ij}$ and the support by $\delta(p)\rightarrow D_{d,d,d,d}$,
\begin{align}\label{eq:grav_final}
    \hexchange{-0.5ex}{1}{2}{3}{4}\,=\frac{1}{8}\frac{1}{\DD_{12}+2d}(\DD_{13}^2+\DD_{23}^2-\DD_{12}^2)D_{d,d,d,d}\,.
\end{align}
This time we need an additional shift in the propagator, $\frac{1}{s_{12}}\rightarrow\frac{1}{\DD_{12}+2d}$. It originates from the coupling between the graviton and background curvature that appears in the equation of motion (see appendix~\ref{sec:graviton}). We note that $2d$ is exactly the eigenvalue of the massless graviton's quadratic Casimir operator. 
%
Thus, the equation above confirms the more general prescription \eqref{propreplacement},
$(p_I^2+m^2)^{-1}\rightarrow (\DD_I+\qc{})^{-1}$, connecting propagators of flat space amplitudes and nonlocal products of conformal generators.
That is, while changing the momentum $p_I^{\mu}$ to the conformal generator $D_I^{AB}$, it is necessary to generalize the mass in the propagator to the eigenvalue of the quadratic Casimir operator. In the next section, we will see that $\qc{\rm graviton}=2d$ plays a crucial role when we convert the differential form of the correlator into a sum of $D$-functions with even boundary dimension --- it ensures that the sum truncates to a finite number of terms.

If the scalars $\varphi_1$ and $\varphi_2$ are of one flavor and  $\tilde\varphi_3$ and $\tilde\varphi_4$ are of another, the graviton can only mediate the s-channel interaction. Thus eq.~\eqref{eq:grav_final} is the complete four-scalar boundary correlator for the Lagrangian~\eqref{eq:gravity2}. We write the result in a form that resembles a double copy~\eqref{eq:scalardc2},
\begin{align}\label{eq:AdSdc1}
    \mathcal{M}(\varphi_1\varphi_2\tilde\varphi_3\tilde\varphi_4)&=\frac{1}{8}\frac{1}{\DD_{12}+2d}(\DD_{13}^2+\DD_{23}^2-\DD_{12}^2)D_{d,d,d,d}\nonumber\\
    &=\frac{1}{16}\frac{1}{\DD_{12}+2d}\Big[(\DD_{23}-\DD_{13})^2+2d \DD_{12}\Big]D_{d,d,d,d}\nonumber\\
    &\quad +\frac{1}{16}\frac{1}{\DD_{23}+2d}(\DD_{23}^2+2d\DD_{23})D_{d,d,d,d}\nonumber\\
    &\quad+\frac{1}{16}\frac{1}{\DD_{13}+2d}(\DD_{13}^2+2d\DD_{13})D_{d,d,d,d}\,.
\end{align}

If all four scalars have the same flavor, the complete correlator is a sum of 
the s-, t- and u-channels, 
\begin{align}\label{eq:AdSdc2}
    \mathcal{M}(\varphi_1\varphi_2\varphi_3\varphi_4) &= \;\hexchange{-0.5ex}{1}{2}{3}{4}\;+\;\hexchange{-0.5ex}{1}{4}{2}{3}\;+\;\hexchange{-0.5ex}{1}{3}{4}{2}\,\\
    &=\frac{1}{16}\frac{1}{\DD_{12}+2d}\Big[(\DD_{23}-\DD_{13})^2+2d \DD_{12}\Big]D_{d,d,d,d}+(\textrm{t and u channel})\,.\nonumber
\end{align}
This bears remarkable similarity with the flat space double copy relation~\eqref{eq:flatdc} for same-flavor scalars. 

Interestingly, both eq.~\eqref{eq:AdSdc1} and~\eqref{eq:AdSdc2} have the general form
\begin{align}
    \mathcal{M}&=\frac{1}{16}\left[\frac{1}{\DD_{12}+2d}(\hat{N}_{\rm s}^2+2d\DD_{12})+\frac{1}{\DD_{23}+2d}(\hat{N}_{\rm t}^2+2d\DD_{23})\right.\nonumber\\
    &\quad\quad\quad\left.+\frac{1}{\DD_{13}+2d}(\hat{N}_{\rm u}^2+2d\DD_{13})\right]D_{d,d,d,d}\,,
    \label{eq:2copy_plus_extra}
\end{align}
where $\hat{N}_{\rm s,t,u}$ are the gauge theory BCJ numerators (with suitable replacements of Mandelstam invariants with differential operators). This equation suggests a generic double copy prescription for scalar correlators on AdS space, in which the straightforward extension of the flat space double copy is further corrected 
by numerator terms proportional to the quadratic Casimir of intermediate states and lower-dimension differential operators.
Since the evidence for this conjectured extension is only eqs.~\eqref{eq:AdSdc1} and~\eqref{eq:AdSdc2}, it may need to be further modified to accommodate data from higher-point AdS scalar boundary correlators and other theories.

We note that, since the action of conformal generators on $D_{d,d,d,d}$ yields a factor of $d$ (as can be seen from e.g. the integral representation of the contact integral), the second term in the gravitational numerator, which causes a departure from the naive double copy and is proportional to the quadratic Casimir in the graviton propagator, is subleading in the large-$d$ limit. In this limit, the usual double copy relation is restored.

\subsection{Local factor of scalar boundary correlators and remarks on color-kinematics duality \label{largeD}}

We showed in the explicit examples discussed above, and in general in section~\ref{BGsection}, that the nonlocal part of AdS boundary correlators, which is in one-to-one correspondence with bulk-bulk propagators, can be extracted as an overall factor despite the superficial noncommutativity of differential operators.
The examples discussed above also show that the remaining local part can be written as a sequence of point-wise conformal generators acting on a single contact integral, in accordance with the expected differential form of the correlators in section~\ref{reviewdiffrep}.
Moreover, we showed that certain AdS boundary correlators in the differential representation are formally very similar to the corresponding flat space amplitudes. 
In this section, we comment on the general form of the local differential operator factor for scalar correlators and explore the depth of its similarity with flat space amplitudes. We will see that the terms with the highest number of conformal generators can always be obtained by a formal replacement of Mandelstam invariants in flat space amplitudes.

For a generic scalar Witten diagram, after we trade the bulk-bulk propagators for the nonlocal operators $({\DD_{I}+\qc{}})^{-1}$, the contact interaction is schematically a linear combination of terms of the form
\begin{align}\label{eq:contact_int}
    \int_{\ads} dX \nabla^{m_1}E_{\Delta_1}\nabla^{m_2}E_{\Delta_2}\ldots\nabla^{m_n}E_{\Delta_n}\,.
\end{align}
where $E_{\Delta}\equiv E_{\Delta}(X,P)=\frac{1}{(-2P\cdot X)^{\Delta}}$ is the shorthand notation for a scalar bulk-boundary propagator. 
Indices are contracted with $G_{AB}$, which are inherited from vertex functions and bulk-bulk propagators. Because of our Berends-Giele-type construction, the equation above implicitly contains $G_{AB}$ which are acted upon by derivatives.
Before explicitly evaluating any of the derivatives, a cubic $n$-point gluon-exchange Witten diagram will contain $\sum_{i=1}^n m_i=n-2$ derivatives, and a graviton exchange diagram will have $\sum_{i=1}^{n}m_i=2n-4$. 

To obtain the differential representation, we need to translate eq.~\eqref{eq:contact_int} into boundary conformal generators $D_i^{AB}$ acting on a single $D$-function. Starting from the identity
\begin{align}\label{identity0}
    \nabla_A E_{\Delta} = G_{A}{}^B(2\Delta P_B E_{\Delta+1}) = 2 \Delta P_{A}E_{\Delta+1}+\frac{\Delta X_A}{X^2}E_{\Delta}
\end{align}
with $\nabla_A$ defined in eq.~\eqref{laplacian}, it is easy to see that
\begin{align}\label{eq:DE}
    \nabla_A E_{\Delta_1}\nabla^A E_{\Delta_2}=\frac{1}{X^2}(D_1\cdot D_2) E_{\Delta_1}E_{\Delta_2}\,.
\end{align}
Since $D_1\cdot D_2$ acts only on the boundary data, we can factor it out of the bulk integral. This identity is of the exact form as the flat space relation 
\begin{align}
(i\partial_{\mu})e^{ip_1x}(i\partial^{\mu})e^{ip_2x}=(p_1\cdot p_2)e^{i(p_1+p_2)x}\,.
\end{align}
Effectively, we have traded two contracted $\nabla$'s acting on $E_{\Delta_1}$ and $E_{\Delta_2}$ for $D_1\cdot D_2$, which makes the contact interaction one step closer to the $D$-function. 

The analogy with flat space still holds for certain cases that have more complicated contractions among $\nabla$'s, 
for example,
\begin{align}\label{eq:DE2}
    \nabla_A\nabla_B E_{\Delta_1}\nabla^A E_{\Delta_2}\nabla^B E_{\Delta_3}=\frac{1}{X^4}(D_1\cdot D_2)(D_1\cdot D_3)E_{\Delta_1}E_{\Delta_2}E_{\Delta_3}\,.
\end{align}
There is no ordering ambiguity between $D_1\cdot D_2$ and $D_1\cdot D_3$ because their commutator annihilates the product of bulk-boundary propagators it acts on. We note that for gluon-exchange correlators up to six points and for graviton exchange correlators up to four points, the contact interaction only involves terms like eq.~\eqref{eq:DE} and~\eqref{eq:DE2}. 
Hence, in these cases the numerator of the boundary correlator can be obtained directly from replacing $p_i\cdot p_j$ by $D_i\cdot D_j$ in the flat space amplitude, as we found in the explicit examples discussed in previous sections.

The exact correspondence breaks, however, for more generic cases, such as when the local factor of the correlator contains
\begin{subequations}
\label{eq:DE3}
\begin{align}
    &\nabla_A\nabla_B E_{\Delta_1}\nabla^A\nabla^B E_{\Delta_2}=\frac{1}{X^2}\Big[(D_1\cdot D_2)^2-(d-1)(D_1\cdot D_2)\Big]E_{\Delta_1}E_{\Delta_2}\,,\label{eq:DE3a}\\
    &\nabla_A\nabla_B\nabla_C E_{\Delta_1}\nabla^A E_{\Delta_2}\nabla^B E_{\Delta_3}\nabla^C E_{\Delta_4}\nonumber\\
    &=\frac{1}{X^6}\Big[(D_1\cdot D_2)(D_1\cdot D_3)(D_1\cdot D_4)-(D_1\cdot D_3)(D_2\cdot D_4)\Big]E_{\Delta_1}E_{\Delta_2}E_{\Delta_3}E_{\Delta_4}\,.
\end{align}
\end{subequations}
These relations can be derived by using the commutators,
\begin{align}
    &[\nabla^A,\nabla^B]= \frac{1}{X^2}D_X^{AB}\,,&& [\nabla^A,D_X^{CD}]=\frac{\eta^{AC}\nabla^D-\eta^{AD}\nabla^C}{\sqrt{2}}\,,\nonumber\\
    &[\nabla_A,G_{BC}]=-\frac{G_{AB}X_C+G_{AC}X_B}{X^2}\,,&& [\nabla_A,X_B]=G_{AB}\,.
    \label{comms}
\end{align}
Such commutators typically reduce the number of derivatives by one, which is the reason that terms with lower powers of $D_i\cdot D_j$ appear in eq.~\eqref{eq:DE3}.

We will prove in section~\ref{sec:proof} that a differential representation of the form~\eqref{eq:diff0} to~\eqref{eq:diffA} exists for all scalar boundary correlators. We will see that the terms of highest degree of $D_i$ are obtained by directly replacing a bulk derivative with AdS symmetry generator acting on the same bulk-boundary propagator. 
Thus, the replacement $p_i\cdot p_j\rightarrow D_i\cdot D_j$ in the flat-space kinematic numerators will always give correctly the terms with the highest number of $D_i\cdot D_j$. Moreover, such terms correspond to the leading order in the large-$d$ limit of the complete local factor of the correlator.\footnote{We assume here that $\Delta_i$ scale as $d^{n>1/2}$ for all the boundary operators. Indeed, the action of $D_i E_\Delta \propto \Delta$, cf. eq.~\eqref{identity0}, $D_i\cdot D_j$ dominates if $\lim_{d\rightarrow \infty}({d}/{\Delta^2}) =0 $.  }
Therefore, we conclude that in the large-$d$ limit the differential representation of the massless scalar AdS boundary correlators exactly agrees with the flat space amplitudes under the replacement $p_i\cdot p_j\rightarrow D_i\cdot D_j$ (or $s_{ij}\rightarrow \DD_{ij}$). 

It also follows that, by starting with flat-space kinematic numerators obeying color-kinematics duality, this replacement yields a large-$d$ differential representation for scalar correlators that obeys color-kinematics duality in AdS space (and also the AdS analog of the fundamental BCJ relations). 
Moreover, since the symmetry generators can be taken to commute in this limit, the double-copy of scalar correlators will yield scalar correlators in the corresponding double-copy theory. Section~\ref{gravexchange} illustrates this general observation: by dropping terms with a single $D_i\cdot D_j$ in eq.~\eqref{eq:2copy_plus_extra} that equation takes the standard flat-space double-copy form.
We will return to this point in section~\ref{sec:conclusion} where we will also briefly discuss various other double-copy prescriptions at finite $d$.

We note that this conclusion is based on a term-by-term analysis of the contact interaction. In a physical theory, which contains a combination of terms of the form~\eqref{eq:contact_int} in the local factor of correlators, it is possible that the terms with fewer $D_i\cdot D_j$ products cancel each other out such that the correspondence to flat space amplitudes goes beyond the large-$d$ limit. Indeed, in the scalar correlator due to graviton exchange, index contractions as in eq.~\eqref{eq:DE3a} can appear for a certain choice of Feynman rules. The subleading terms in the large-$d$ limit will cancel at the end, and the resulting correlator is related to the flat space amplitude by the direct replacement $s_{ij}\rightarrow\DD_{ij}$.

\section{Differential representation for scalar correlators}\label{sec:proof}

In this section, we show that a differential representation exists for AdS scalar boundary correlators in any AdS field theory. Given the separation into nonlocal and local parts and the form of the nonlocal part discussed in section~\ref{BGsection} we only need to show that the local part of the correlators can be expressed in terms of generators of the AdS symmetry group acting on a single $D$-function. 
Starting from the examples considered in section~\ref{largeD}, let us consider a generic AdS scalar constructed from $2n$ covariant derivatives acting on some number of scalar bulk-boundary propagators,
\begin{equation}
\label{generalterm}
\begin{split}
    \mathcal{F}_{2n}&\equiv T^{A_1\ldots A_jB_1\ldots B_k\ldots C_1\ldots C_l}(\nabla_{A_1}\ldots\nabla_{A_j}E_{\Delta_1})(\nabla_{B_1}\ldots\nabla_{B_k}E_{\Delta_2})\\
    &\quad \times \ldots(\nabla_{C_1}\ldots\nabla_{C_l}E_{\Delta_m})\,,
    \end{split}
\end{equation}
where the tensor $T$ is a monomial of degree $n$ in $\eta^{AB}$ (so it has $2n$ indices) that contracts all indices of the derivatives.\footnote{
We assume that the tensor $T$ does not contract the indices of two derivatives acting on the same bulk-boundary propagator. If such a contraction occurs, we can always commute one of the derivatives to the front and move it on the other factors through integration-by-parts, while generating additional terms with fewer derivatives due to the nonzero commutators. Alternatively, we can also move both derivatives next to the propagator (while again generating terms with fewer derivatives due to the nonzero commutators) and use the equation of motion to trade the resulting AdS Laplacian for the AdS mass parameter.} 

%
%

Based on the examples in section~\ref{largeD}, we would like to prove that $\mathcal{F}_{2n}$ has the manifestly local form of polynomial in $D_i^{AB}$ acting on (the integrand of) a single $D$-function. More specifically, 
\begin{align}\label{eq:F2n}
    \mathcal{F}_{2n}=\frac{1}{X^2}\Big[M_{2n}(D_i)+P_{2n-2}(D_i)+\ldots+P_2(D_i)+P_{0}\Big] \prod_{i}E_{\Delta_i}\,,
\end{align}
where $M_{2n}$ is a monomial of degree $2n$ in $D_i^{AB}$, and $P_{2k\leqslant 2n-2}$ are polynomials of degree $2k$ in $D_i$. 
Compared with the leading monomial $M_{2n}$, their degree decreases by at least two units. The readers not interested in the details of the proof may skip the rest of this section.
We will show that $M_{2n}$ can be obtained from the corresponding flat-space expression via the replacement $p_{i}\cdot p_{j}\mapsto D_{i}\cdot D_{j}$ and that the sub-leading terms, $P_{2n-2i}$, are suppressed in the large-$d$ and $\Delta$ limit.

We will prove the statement recursively, with eq.~\eqref{eq:F2n} serving as the recursive assumption. The starting point is eq.~\eqref{eq:DE}, which is of the form~\eqref{eq:F2n} with the monomial $M_2=D_1\cdot D_2$ and can be obtained through the replacement $p_1\cdot p_2\rightarrow D_1\cdot D_2$ in the numerator of the corresponding flat space expression in which the covariant derivatives are simply partial derivatives and $E_\Delta \rightarrow e^{ipx}$.
We then consider eq.~\eqref{generalterm} with two additional derivatives, which without loss of generality, we choose to act on $E_{\Delta_1}$ and $E_{\Delta_2}$, and define
\begin{align}
    \mathcal{F}_{2n+2}\equiv T^{A_1\ldots A_jB_1\ldots B_kC_1\ldots C_l}(\nabla_{A_1}\ldots \nabla_N\ldots\nabla_{A_j}E_{\Delta_1})(\nabla_{B_1}\ldots\nabla^N\ldots\nabla_{B_k}E_{\Delta_2})\ldots\, ,
\end{align}
where the ellipsis stand for additional factors present in eq.~\eqref{generalterm} that 
do not depend on $E_{\Delta_1}$ and $E_{\Delta_2}$.
The goal is to show that $\mathcal{F}_{2n+2}$ has the same form~\eqref{eq:F2n} with $n\rightarrow n+1$, i.e. that it begins with a degree-$(2n+2)$ monomial in $D_i$ and continues with subleading polynomials whose degrees decrease by two units. 

To evaluate $\mathcal{F}_{2n+2}$ we start by commuting $\nabla_N$, using eq.~\eqref{comms}, to act directly on $E_{\Delta_1}$ and $E_{\Delta_2}$, 
\begin{align}\label{eq:larged0}
    &\nabla_{A_1}\ldots \nabla_N\ldots\nabla_{A_j}E_{\Delta_1}=\nabla_{A_1}\ldots\nabla_{A_j}\nabla_NE_{\Delta_1}+\mathcal{O}(D_{X}\nabla^{j-1}E_{\Delta_1})\,,\nonumber\\
    &\nabla_{B_1}\ldots\nabla^{N}\ldots\nabla_{B_k}E_{\Delta_2}=\nabla_{B_1}\ldots\nabla_{B_k}\nabla^NE_{\Delta_2}+\mathcal{O}(D_{X}\nabla^{k-1}E_{\Delta_2})\,,
\end{align}
where $\mathcal{O}(D_X\nabla^{\ldots})$ are terms resulting from the commutator $[\nabla^A,\nabla^B]=-D_X^{AB}$. They contain at least one fewer derivative than the 
left hand side. In the product, the cross term $\nabla_{B_1}\ldots\nabla_{B_k}\nabla^NE_{\Delta_2}\mathcal{O}(D_X\nabla^{j-1}E_{\Delta_1})
$ appears to have only one fewer derivative than the leading term, with the term with 
the largest number of derivatives containing a factor of the form
\begin{align}\label{lastX}
    &\nabla_{A_1}\ldots\nabla_{A_{i-1}}D_X^{NA_i}\nabla_{A_{i+1}}\ldots\nabla_{A_j}E_{\Delta_1}\nabla_{B_1}\ldots\nabla_{B_k}\nabla_NE_{\Delta_2}\\
    &=\frac{X^{A_i}}{X^2}D_1\cdot D_2 \nabla_{A_1}\ldots\nabla_{A_{i-1}}\nabla_{A_{i+1}}\ldots\nabla_{A_j}E_{\Delta_1}\nabla_{B_1}\ldots\nabla_{B_k}E_{\Delta_2}+(\text{commutators})\,, \nonumber
\end{align}
where the term written comes from the explicit evaluation of $D_X^{NA_i}$ and $\nabla_N$. The commutators $[D_X,\nabla]$ and $[X,\nabla]$, arising from pushing $D_X$ to $E_{\Delta_1}$ and from extracting $X^{A_i}$, lead to lower number of derivatives.
Since all indices are contracted, $X^{A_i}$ will be contracted with another $\nabla_{A_i}$. Because $X^{A_i}\nabla_{A_i}=0$ , eq.~\eqref{lastX} can contribute to the complete product only the commutator of $X^{A_i}$ with another derivative, $[\nabla^A,X^B]=G^{AB}$, which further reduces the number of derivatives by one. It therefore follows that the cross-terms $\nabla_{B_1}\ldots\nabla_{B_k}\nabla^ME_{\Delta_2}
\mathcal{O}(D_X\nabla^{j-1}E_{\Delta_1})$ also contribute terms with two derivatives fewer than the initial number. That is, 
\begin{align}
    &(\nabla_{A_1}\ldots \nabla_M\ldots\nabla_{A_j}E_{\Delta_1})(\nabla_{B_1}\ldots\nabla^M\ldots\nabla_{B_k}E_{\Delta_2})\\
    &=(\nabla_{A_1}\ldots\nabla_{A_j}\nabla_ME_{\Delta_1})(\nabla_{B_1}\ldots\nabla_{B_k}\nabla^ME_{\Delta_2})+(\text{subleading)}\,,\nonumber
\end{align}
where the subleading terms have the form
\begin{align}\label{eq:sb0}
  (\text{subleading)} \sim  \mathcal{O}(D_1\cdot D_2\nabla^{2n-2}E_{\Delta_1}E_{\Delta_2}\ldots)\,,
\end{align}
i.e., they have at least two fewer derivatives, as we expected from eq.~\eqref{eq:F2n}. We note that the lower order terms contain more generic contractions between $\nabla^A$ and two $D_i^{AB}$ acting on different bulk-boundary propagators. Using the same relations, we can show that the number of derivatives is even and at most $2n-2$.

Having seen that the difference between the highest and next-to-highest number of derivatives in ${\cal F}_{2n+2}$ is two units, we proceed to show that ${\cal F}_{2n+2}$ can be written in terms of the AdS symmetry generators.
To this end, we evaluate $\nabla^M E_{\Delta}$ using eq.~\eqref{identity0}, and collect the result as
\begin{align}\label{eq:larged1}
    &(\nabla_{A_1}\ldots\nabla_{A_j}\nabla_NE_{\Delta_1})(\nabla_{B_1}\ldots\nabla_{B_k}\nabla^NE_{\Delta_2})\nonumber\\
    &=4\Delta_1\Delta_2(P_1\cdot P_2)\nabla_{A_1}\ldots\nabla_{A_j}E_{\Delta_1+1}\nabla_{B_1}\ldots\nabla_{B_k}E_{\Delta_2+1} \nonumber\\
    &\quad -\Delta_1\Delta_2\nabla_{A_1}\ldots\nabla_{A_j}E_{\Delta_1}\nabla_{B_1}\ldots\nabla_{B_k}E_{\Delta_2}+(\text{subleading})\,.
\end{align}
The subleading terms contain various commutators $[\nabla^A,X^B]=G^{AB}$ and thus have fewer derivatives. Those with the most derivatives have the form
\begin{align}\label{eq:sub1}
    2\Delta_1\Delta_2 P_{1,N}G^{N}{}_{B_i}\nabla_{A_1}\ldots\nabla_{A_j}E_{\Delta_1+1}\nabla_{B_1}\ldots\nabla_{B_{i-1}}\nabla_{B_{i+1}}\ldots\nabla_{B_k}E_{\Delta_2}\,.
\end{align}
Since the index $B_i$ is contracted with a derivative acting on another bulk-boundary propagator $E_{\Delta_\ell}$, we can use the identity\footnote{In principle, there are additional derivatives acting on $E_{\Delta_1+1}$ and $E_{\Delta_\ell}$ in eq.~\eqref{eq:sub1}. To take them into account, we need to add terms containing a commutator $[\nabla^A,G^{BC}]$ to eq.~\eqref{eq:sub2}, similar to eq.~\eqref{lastX}. They all contain fewer derivatives so we omit them from our discussion.}
\begin{align}\label{eq:sub2}
    2\Delta_1 P_{1,N}G^{N}{}_{B_i}E_{\Delta_1+1}\nabla^{B_i}E_{\Delta_\ell}=\frac{1}{X^2}D_{1}\cdot D_\ell E_{\Delta_1}E_{\Delta_\ell}\,,
\end{align}
to trade one derivative on $E_{\Delta_\ell}$ for $D_{\ell}$, and the $P_1$ factor for $D_1$. Therefore, the terms with the most derivatives in the subleading term in eq.~\eqref{eq:larged1} have the form
\begin{align}\label{eq:sb1}
    \mathcal{O}(\Delta_2 D_1\cdot D_{\ell}\nabla^{2n-2}E_{\Delta_1}E_{\Delta_2}\ldots)\,,
\end{align}
plus the $1\leftrightarrow 2$ exchange. Now we consider
\begin{align}\label{eq:larged2}
    &\frac{1}{X^2}D_1\cdot D_2(\nabla_{A_1}\ldots\nabla_{A_j}E_{\Delta_1})(\nabla_{B_1}\ldots\nabla_{B_k}E_{\Delta_2})\nonumber\\
    &=4\Delta_1\Delta_2(P_1\cdot P_2)\nabla_{A_1}\ldots\nabla_{A_j}E_{\Delta_1+1}\nabla_{B_1}\ldots\nabla_{B_k}E_{\Delta_2+1} \nonumber\\
    &\quad -\Delta_1\Delta_2\nabla_{A_1}\ldots\nabla_{A_j}E_{\Delta_1}\nabla_{B_1}\ldots\nabla_{B_k}E_{\Delta_2}+(\text{subleading})\,.
\end{align}
The first two terms agree with the first two terms in eq.~\eqref{eq:larged1}. For the subleading terms in eq.~\eqref{eq:larged2}, we again focus on those with the most derivatives. They are of two types: one is of the form~\eqref{eq:sb1}, the other one comes from
\begin{align}
    \Delta_1\Delta_2(P_1\cdot P_2)G_{A_i}{}^{B_m}\nabla_{A_1}..\nabla_{A_{i-1}}\nabla_{A_{i+1}}..\nabla_{A_j}E_{\Delta_1+1}\nabla_{B_1}..\nabla_{B_{m-1}}\nabla_{B_{m+1}}..\nabla_{B_k}E_{\Delta_2+1}\,.
\end{align}
Up to terms with less derivative, we can trade $\Delta_1\Delta_2 P_1\cdot P_2$ with $D_1\cdot D_2$. The $G_{A_i}{}^{B_m}$ factor can bring at most a factor of $d$ when the index $A_i$ and $B_m$ are contracted with each other. Therefore, the terms of this second type can be formally written as
\begin{align}\label{eq:sb2}
    \mathcal{O}(d\,D_1\cdot D_2 \nabla^{2n-2}E_{\Delta_1}E_{\Delta_2}\ldots)\, ,
\end{align}
i.e., they can be written in terms of $D_1\cdot D_2$ acting on AdS scalars with fewer derivatives which, following our recursion, have the form \eqref{eq:F2n}.

Applying the argument above to the subleading terms in eqs.~\eqref{eq:sb1} and~\eqref{eq:sb2} implies that they contain at most $2n-2$ derivatives composed of $D_i\cdot D_j$ and $\nabla^A$. The number of derivatives drops only in steps of two, accompanied by a coefficient at most proportional to $\Delta$ or $d$. 

Combining this with eq.~\eqref{eq:larged0}, \eqref{eq:larged1} and~\eqref{eq:larged2}, we find that
\begin{align}\label{eq:F2}
    \mathcal{F}_{2n+2}&=\frac{1}{X^2}(D_1\cdot D_2)\mathcal{F}_{2n}+(\text{subleading})\nonumber\\
    &=\frac{1}{X^2}\Big[M_{2n+2}(D_i)+\tilde{P}_{2n}(D_i)+\ldots+\tilde{P}_{0}\Big] \prod_{i}E_{\Delta_i}\, .
\end{align}
It also follows that the leading monomial $M_{2n+2}(D_i)$ is related to the leading one $M_{2n}(D_i)$ of $\mathcal{F}_{2n}$ as
\begin{align}
    M_{2n+2}(D_i)=(D_1\cdot D_2) M_{2n}(D_i)\,.
\end{align}
Therefore, the leading monomial can be obtained from the corresponding flat space amplitude numerator by the replacement $p_i\cdot p_j\rightarrow D_i\cdot D_j$. Note that the ordering of $D_i\cdot D_j$ only affects the subleading terms. 
The subleading polynomial $\tilde{P}_{2n}(D_i)$ receives contributions from the terms in eqs.~\eqref{eq:sb0}, \eqref{eq:sb1} and~\eqref{eq:sb2}. They all have a factor of overall $D_i\cdot D_j$ and $2n-2$ covariant derivatives $\nabla$ contracted among themselves. Then according to the recursive assumption~\eqref{eq:F2n}, they can be converted to a polynomial of degree $2n$ in $D_i$, with coefficients that are at most linear in $d$ or in $\Delta_i$. In addition, the generators $D_i$ appear only in the product $D_i\cdot D_j$. 

We note that the polynomials of degree $2n-2$ and lower in eq.~\eqref{eq:F2} receive contributions from more complicated contractions between $\nabla^A$ and at most two $D_i^{AB}$.
To offer another perspective on the argument above that two $D_i^{AB}$, if they appear, will always contract with each other and factor out of the bulk integral, allowing us to use the recursive assumption~\eqref{eq:F2n} to write $\mathcal{F}_{2n+2}$ completely in terms of polynomials of $D_i\cdot D_j$, let us consider another possible index contraction --- $\nabla^A$ contracted with $D_1^{AB}$. Then, the identity 
\begin{align}
    D_1^{AB}E_{\Delta_1}\nabla_A E_{\Delta_\ell}=\frac{X^B}{X^2}D_{1}\cdot D_\ell E_{\Delta_1}E_{\Delta_\ell}
\end{align}
converts it into $D_1\cdot D_\ell$ and $X^B$. If the free index $B$ is contracted with another derivative $\nabla$, then the $X^B$ factor will effectively reduce the number of derivatives because $X^A\nabla_A = 0$. If it is contracted with the other $D^{AB}$, then we use
\begin{align}
\frac{\sqrt{2}\,X^B}{X^2}D_{AB}E_{\Delta}=\nabla_A E_{\Delta}
\end{align}
to convert it into a $\nabla$. For both cases, additional $\nabla$'s acting on $E_{\Delta_1}$ and $E_{\Delta_\ell}$ will lead to terms with fewer derivatives by a step of two through commutators. 
We can then proceed to lower order terms and systematically separate $D_i\cdot D_j$ from $\nabla$'s. With the recursive assumption~\eqref{eq:F2n}, this completes the proof that $\mathcal{F}_{2n+2}$, as given in eq.~\eqref{eq:F2}, has the same form as eq.~\eqref{eq:F2n}. Importantly, as stated in section~\ref{largeD}, the polynomials $\tilde{P}_{2k\leqslant 2n}$ are subleading in the large-$d$ limit compared with $M_{2n+2}(D_i)$.
%

\section{Evaluating the Differential Representation 
\label{sec: evaluationofdiffrep}}

We have seen that scalar correlators in differential representation have very simple expressions that resemble flat space amplitudes. Now we must address the question of how we should evaluate the Witten diagrams. For simplicity, we first assume that the conformal weights for boundary and bulk states are chosen such that the position-space correlators can be expressed in terms of a finite sum of $D$-functions. To evaluate the Witten diagrams, one can of course always go back to the conventional position-space representation, for example, eq.~\eqref{eq:gluon_exchange}, for which various techniques have been developed in the past. In contrast, we will show that the expansion of the correlators in terms of $D$-functions can be derived by only utilizing the action of the operator $\DD_{ij}$ on products of 
\begin{align}
    P_{ij}\equiv -2P_i\cdot P_j
\end{align}
and $D$-functions. Our method is independent of the explicit representation and gauge fixing of the bulk-bulk propagators, which significantly simplifies the computation compared with existing methods.

\subsection{Scalar Exchange}

We begin with evaluating the scalar exchange diagram~\eqref{eq:sexchange}. Our strategy will be to build a recursion relation for the quantity
\begin{align}
    \frac{1}{\DD_{12}+\Delta(\Delta-d)}\Big[ P_{12}^{-k+1}D_{\Delta_1-k+1,\Delta_2-k+1,\Delta_3,\Delta_4}\Big]\,.
\end{align}
Once the recursion is solved, we will need only the $k=1$ term.
%
To derive the recursion, we start with the action of $\DD_{12}+\Delta(\Delta-d)$ on $P_{12}^{-k}D_{\Delta_1-k,\Delta_2-k,\Delta_3,\Delta_4}$,
\begin{align}
    &\Big[\DD_{12}+\Delta(\Delta-d)\Big]\Big[P_{12}^{-k}D_{\Delta_1-k,\Delta_2-k,\Delta_3,\Delta_4}\Big]\\
    &=4(\Delta_1-k)(\Delta_2-k)P_{12}^{-k+1}D_{\Delta_1-k+1,\Delta_2-k+1,\Delta_3,\Delta_4}\nonumber\\
    &\quad-(\Delta_1+\Delta_2+\Delta-d-2k)(\Delta_1+\Delta_2-\Delta-2k)P_{12}^{-k}D_{\Delta_1+k,\Delta_2+k,\Delta_3,\Delta_4}\,.\nonumber
\end{align}
We then move the operator to the right hand side. After some rearrangements, we obtain the desired recursion:
\begin{align}
&\frac{1}{\DD_{12}+\Delta(\Delta-d)}\Big[P_{12}^{-k+1}D_{\Delta_1-k+1,\Delta_2-k+1,\Delta_3,\Delta_4}\Big]=\frac{P_{12}^{-k}D_{\Delta_1-k,\Delta_2-k,\Delta_3,\Delta_4}}{4(\Delta_1-k)(\Delta_2-k)}\\
&+\frac{(\Delta_1{+}\Delta_2{+}\Delta{-}d{-}2k)(\Delta_1{+}\Delta_2{-}\Delta{-}2k)}{4(\Delta_1-k)(\Delta_2-k)}\frac{1}{\DD_{12}+\Delta(\Delta-d)}\Big[P_{12}^{-k}D_{\Delta_1-k,\Delta_2-k,\Delta_3,\Delta_4}\Big]\,.\nonumber
\end{align}
It is interesting to note that the recursion terminates if either $k_{\rm max}=\frac{\Delta_1+\Delta_2-\Delta}{2}$ or $k_{\rm max}=\frac{\Delta_1+\Delta_2+\Delta-d}{2}$ is an integer.\footnote{If both are integers, then the recursion terminates at $k_{\rm max}=\frac{\Delta_1+\Delta_2-\Delta}{2}$ since it is the smaller one of the two if we assume $\Delta$ is the larger root of $\Delta(\Delta-d)=m^2$.} For such cases, eq.~\eqref{eq:sexchange} can be written as a finite sum of $D$-functions,
\begin{align}\label{eq:scalar_expand}
    &\frac{1}{\DD_{12}+\Delta(\Delta-d)}D_{\Delta_1,\Delta_2,\Delta_3,\Delta_4}=\sum_{l=1}^{k_{\rm max}}\frac{(\Delta_1)_{-l}(\Delta_2)_{-l}P_{12}^{-l}D_{\Delta_1-l,\Delta_2-l,\Delta_3,\Delta_4}}{4\!\left(\frac{\Delta_1+\Delta_2-\Delta}{2}\right)_{1-l}\left(\frac{\Delta_1+\Delta_2+\Delta-d}{2}\right)_{1-l}}\,,
\end{align}
which is exactly the $D$-function expansion for this Witten diagram~\cite{DHoker:1999mqo}. As before, $(a)_n\equiv\frac{\Gamma(a+n)}{\Gamma(a)}$ is the Pochhammer symbol.

We have explicitly restricted ourselves to correlators that can be represented as finite sums of $D$-functions. Such expressions manifestly obey desirable boundary conditions for the correlator. 
 Correlators that cannot be expressed as finite sums of $D$-functions will generically depend on multiple infinite families $D$-functions, see Appendix~C of Ref.~\cite{Zhou:2018sfz} for an explicit example. 

\subsection{Gluon Exchange}

The scalar exchange computation demonstrates the general strategy for the evaluation of Witten diagrams in differential representation. For the gluon exchange diagram~\eqref{eq:sgauge}, we first expand the numerator in terms of $D$-functions,
\begin{align}\label{eq:gexchange3}
    \frac{1}{\DD_{12}}(\DD_{23}-\DD_{13})D_{d,d,d,d}=\frac{4d^2}{\DD_{12}}(P_{23}D_{d,d+1,d+1,d}-P_{13}D_{d+1,d,d+1,d})\,.
\end{align}
Then the action of $\DD_{12}^{-1}$ can be obtained by a similar recursion. In fact, we only need to work out its action on $P_{13}D_{d+1,d,d+1,d}$ since the other can be simply obtained by relabelling. We start with
\begin{align}
    \DD_{12}\Big[P_{12}^{-k}P_{13}&D_{d-k+1,d-k,d+1,d}\Big]= 4(d-k)(d-k+1)P_{12}^{-k+1}P_{13}D_{d-k+2,d-k+1,d+1,d}\nonumber\\
    & \quad +2(d-k)P_{12}^{-k}\Big[P_{23}D_{d-k,d-k+1,d+1,d}-P_{12}D_{d-k+1,d-k+1,d,d}\Big]\nonumber\\
    & \quad -2(d-k)(d-2k+1)P_{12}^{-k}P_{13}D_{d-k+1,d-k,d+1,d}\,.
\end{align}
We can eliminate the term $P_{23}D_{d-k,d-k+1,d+1,d}$ on the right-hand side by the identity
\begin{align}
    P_{23}D_{d-k,d-k+1,d+1,d}&=-\frac{d-k}{d}P_{12}D_{d-k+1,d-k+1,d,d}-P_{13}D_{d-k+1,d-k,d+1,d}\nonumber\\
    &\quad -\frac{2k-3d}{2d}D_{d-k,d-k,d,d}\,,
\end{align}
which can be derived from eq.~\eqref{eq:cwi4}. This gives us the desired recursion
\begin{align}
    &\frac{1}{\DD_{12}}\Big[P_{12}^{-k+1}P_{13}D_{d-k+2,d-k+1,d+1,d}\Big]= \frac{1}{4(d{-}k)(d{-}k{+}1)}P_{12}^{-k}P_{13}D_{d-k+1,d-k,d+1,d}\nonumber\\
    & +\frac{d{-}2k{+}2}{2(d{-}k{+}1)}\frac{1}{\DD_{12}}\Big[P_{12}^{-k}P_{13}D_{d-k+1,d-k,d+1,d}\Big]{+}\frac{2d{-}k}{2d(d{-}k{+}1)}\frac{1}{\DD_{12}}\Big[P_{12}^{-k+1}D_{d-k+1,d-k+1,d,d}\Big]\nonumber\\
    &  -\frac{3d-2k}{4d(d-k+1)}\frac{1}{\DD_{12}}\Big[P_{12}^{-k}D_{d-k,d-k,d,d}\Big]\,.
\end{align}
The coefficient of the second term on the right-hand side guarantees that the recursion terminates at $k_{\rm max}=\frac{d}{2}+1$ if $d$ is even. For these cases, the solution is
\begin{align}
    \frac{1}{\DD_{12}}P_{13}D_{d+1,d,d+1,d}&=P_{13}\frac{1}{\DD_{12}{-}d{+}1}D_{d+1,d,d+1,d}\nonumber\\
    &\quad +\sum_{k=1}^{d/2+1}\frac{(d{+}1)_{-k}(2d{-}k)}{2d(d/2{+}1)_{1-k}}\frac{1}{\DD_{12}}P_{12}^{-k+1}D_{d+1-k,d+1-k,d,d}\nonumber\\
    &\quad  -\sum_{k=1}^{d/2+1}\frac{(d{+}1)_{-k}(3d{-}2k)}{4d(d/2{+}1)_{1-k}}\frac{1}{\DD_{12}}P_{12}^{-k}D_{d-k,d-k,d,d}\,.
\end{align}
We can pull the $\DD_{12}^{-1}$ in the last two terms out of the summation. Then using eq.~\eqref{eq:scalar_expand}, we see that they sum into a scalar-exchange diagram with internal weight $\Delta=d-2$ and a contact diagram,
\begin{align}
\frac{1}{\DD_{12}}\left[-\frac{(d-2)(d-1)^2}{d^2}\frac{1}{\DD_{12}-2(d-2)}D_{d,d,d,d}+\frac{2d-1}{2d^2}D_{d,d,d,d}\right]\,.
\end{align}
After partial fractioning the product of inverse operators, we obtain the final result
\begin{align}
    \frac{1}{\DD_{12}}\Big[P_{13}D_{d+1,d,d+1,d}\Big]&=P_{13}\frac{1}{\DD_{12}-d+1}D_{d+1,d,d+1,d}\nonumber\\
    &\quad - \frac{(d-1)^2}{2d^2}\frac{1}{\DD_{12}-2(d-2)}D_{d,d,d,d}+\frac{1}{2}\frac{1}{\DD_{12}}D_{d,d,d,d}\,.
\end{align}
Replacing this into eq.~\eqref{eq:gexchange3}, we get
\begin{align}
    \gexchange{-0.5ex}{1}{2}{3}{4}\;&=\frac{1}{\DD_{12}}(\DD_{23}-\DD_{13})D_{d,d,d,d}\nonumber\\[-1em]
    &=P_{23}\frac{4d^2}{\DD_{12}-d+1}D_{d,d+1,d+1,d}-P_{13}\frac{4d^2}{\DD_{12}-d+1}D_{d+1,d,d+1,d}\, ,
\end{align}
Then, the explicit $D$-function representation of the gluon exchange diagram can be obtained by using eq.~\eqref{eq:scalar_expand} in the second line of the equation above.

\subsection{Graviton Exchange}

Among the three operators in the graviton exchange diagram~\eqref{eq:grav_final}, the last one, $\DD_{12}^2$, commutes with the propagator. It can thus be directly reduced to a scalar exchange diagram and a contact term. For the first two terms, we start by carrying out the derivatives,
\begin{align}\label{eq:graviton_start}
    \frac{1}{8}(\DD_{23}^2+\DD_{13}^2)D_{d,d,d,d}&=2d^2(d{+}1)^2(P_{23}^2D_{d,d+2,d+2,d}+P_{13}^2 D_{d+2,d,d+2,d})\\
    &\quad +\frac{2d^2+3d+2}{4}\DD_{12}D_{d,d,d,d}-d^2(d+1)(d+2)D_{d,d,d,d}\,.\nonumber
\end{align}
The two terms in the second line commute with the propagator so that they become a scalar exchange contribution and a contact term. We are thus left to evaluate
\begin{align}
    \frac{1}{\DD_{12}+2d}(P_{23}^2D_{d,d+2,d+2,d}+P_{13}^2D_{d+2,d,d+2,d})\,.
\end{align}
We again need to derive a recursive relation from
\begin{align}
    (\DD_{12}+2d)\Big[P_{12}^{-k}(P_{23}^2D_{d-k,d-k+2,d+2,d}+P_{13}^2D_{d-k+2,d-k,d+2,d})\Big]\,.
\end{align}
The action of $\DD_{12}$ will generate terms that contain $P_{13}P_{23}$. Using eq.~\eqref{eq:cwi4}, the differential operators should be eliminated in favor of $P_{12}P_{13}$ and $P_{12}P_{23}$ while maintaining the symmetry under $1\leftrightarrow 2$ exchange. 
The final form of the recursion is
\begin{align}\label{eq:gr_recursion}
    &\frac{1}{\DD_{12}+2d}P_{12}^{-k+1}\Big[P_{23}^2D_{d-k+1,d-k+3,d+2,d}+P_{13}^2D_{d-k+3,d-k+1,d+2,d}\Big]\nonumber\\
    &=\frac{1}{4(d-k)(d-k+2)}P_{12}^{-k}\Big[P_{23}^2 D_{d-k,d-k+2,d+2,d}+P_{13}^2 D_{d-k+2,d-k,d+2,d}\Big]\nonumber\\
    &\quad +\frac{(d-k+1)(d-2k+2)}{2(d-k)(d-k+2)}\frac{1}{\DD_{12}{+}2d}P_{12}^{-k}\Big[P_{23}^2 D_{d-k,d-k+2,d+2,d}+P_{13}^2 D_{d-k+2,d-k,d+2,d}\Big]\nonumber\\
    &\quad-\frac{(3d-2k)(3d-2k+2)(d-k+1)}{4d(d+1)(d-k)(d-k+2)}\frac{1}{\DD_{12}+2d}P_{12}^{-k}D_{d-k,d-k,d,d}\nonumber\\
    &\quad +\frac{(3d-2k+2)(3d-2k+3)}{2d(d+1)(d-k+2)}\frac{1}{\DD_{12}+2d}P_{12}^{-k+1}D_{d-k+1,d-k+1,d,d}\nonumber\\
    &\quad -\frac{(d-k+1)(2d-k+2)}{d(d+1)(d-k+2)}\frac{1}{\DD_{12}+2d}P_{12}^{-k+2}D_{d-k+2,d-k+2,d,d}\,,
\end{align}
which terminates at $k_{\rm max}=\frac{d}{2}+1$ for even $d$. We note that the $2d$ shift in the propagator is crucial for this to happen. Should it take a different value, the recursion would not truncate for even $d$. The solution to the above recursion is
\begin{align}\label{eq:gr_step2}
    &\frac{1}{\DD_{12}+2d}\Big[P_{23}^2D_{d,d+2,d+2,d}+P_{13}^2 D_{d+2,d,d+2,d}\Big]\nonumber\\
    &=P_{23}^2 \frac{1}{\DD_{12}}D_{d,d+2,d+2,d}+P_{13}^2 \frac{1}{\DD_{12}}D_{d+2,d,d+2,d}-\frac{d(d{-}1)}{2(d{+}1)^2}\frac{1}{\DD_{12}{-}2d{+}4}D_{d,d,d,d}\nonumber\\
    &\quad -\frac{2d+1}{4d^2(d+1)^2}\frac{1}{\DD_{12}+2d}\DD_{12}D_{d,d,d,d}+\frac{d^2{+}4d{+}2}{2(d+1)^2}\frac{1}{\DD_{12}+2d}D_{d,d,d,d}\,.
\end{align}
Together with eq.~\eqref{eq:graviton_start}, we can now write down the result for the s-channel graviton-exchange contribution to the four-scalar boundary correlator:
\begin{align}
    \hexchange{-0.5ex}{1}{2}{3}{4}&=\frac{1}{8}\frac{1}{\DD_{12}+2d}(\DD_{13}^2+\DD_{23}^2-\DD_{12}^2)D_{d,d,d,d}\nonumber\\[-1em]
    &=2d^2(d+1)^2\left[P_{23}^2\frac{1}{\DD_{12}}D_{d,d+2,d+2,d}+P_{13}^2\frac{1}{\DD_{12}}D_{d+2,d,d+2,d}\right]\nonumber\\
    &\quad-\frac{d^3(d-1)}{\DD_{12}-2d+4}D_{d,d,d,d} -\frac{d^2}{2}P_{12}D_{d+1,d+1,d,d}+\frac{3d^2}{4}D_{d,d,d,d}\,.
\end{align}
In particular, for $d=4$, it becomes
\begin{align}
    \hexchange{-0.5ex}{1}{2}{3}{4}\;\Bigg|_{d=4} &= 800P_{23}^2 \frac{1}{\DD_{12}}D_{4,6,6,4}+800P_{13}^2\frac{1}{\DD_{12}}D_{6,4,6,4}-192\frac{1}{\DD_{12}-4}D_{4,4,4,4}\nonumber\\[-1em]
    &\quad - 8P_{12}D_{5,5,4,4}+12D_{4,4,4,4}\,.
\end{align}
Using eq.~\eqref{eq:scalar_expand} to expand every term into $D$-functions, we find that it indeed agrees with~\cite{DHoker:1999kzh} up to an overall normalization factor.

\subsection{Constraining the graviton exchange correlators}

Very interestingly, we can also use the recursion-based technique described above to construct boundary correlators from a bootstrap point of view. We illustrate this idea by considering the graviton exchange. Our starting point is based on the following requirements:
\begin{enumerate}[label=(R.\arabic*),leftmargin=1.6\parindent]
    \item The scalar AdS correlator with $\Delta_i=d$ can be expressed in terms of a finite sum of $D$-functions in the position space for even $d$.\label{r1}
    \item Every Witten diagram can be written as a set of $\frac{1}{\DD_I+\#}$ acting on a contact interaction, where $\DD_I$'s are in one-to-one correspondence with bulk-bulk propagators in this Witten diagram.\label{r2}
    \item The contact interaction in the previous requirement comes from pinching internal legs of the Witten diagram. Specifically for the graviton exchange, we assume that the minimal coupling to a massless spin-2 particle $h_{\mu\nu}\nabla^{\mu}\varphi\nabla^{\nu}\varphi$ is part of the interaction.\label{r3}
\end{enumerate}
The requirement~\ref{r1} is motivated by explicit examples, such as the ones described in previous sections as well as in certain supersymmetric theories as discussed in e.g.~\cite{DHoker:1999kzh, Rastelli:2017udc, Rastelli:2019gtj}, where boundary correlators are given by a finite sum of $D$-functions. 
The requirement~\ref{r2} is motivated by an analogy with flat space amplitudes. Remarkably, we do not need to fix a priori the shift $\#$ to be the quadratic Casimir.
The requirement~\ref{r3} comes from our expectations regarding the interaction Lagrangian. We only write down the most essential interaction, but keep an open mind that their consistency with~\ref{r1} and~\ref{r2} may lead to additional contributions. We note that the Lagrangian~\eqref{eq:gravity2} also contains a coupling to the trace of the graviton, $h^{\nu}{}_{\nu}\nabla^{\mu}\varphi\nabla_{\mu}\varphi$. We can of course add this term to our requirement, but it will be interesting to see how far we can go without it. Although based on slightly different theories, our approach agrees in spirit with that in Ref.~\cite{Rastelli:2017udc}.

Using only the information above, we now try to construct the s-channel diagram~\eqref{eq:grav_final}. From the interaction in~\ref{r3}, we know that the contact interaction in~\ref{r2} should take the form $\nabla_A E_1\nabla^AE_3\nabla_BE_2\nabla^BE_4$ together with its permutations. In particular, the contraction should be between legs at the opposite ends of the bulk-bulk propagator. 
Therefore, after taking into account Bose symmetry, the bulk integration of the contact interaction gives,
\begin{align}\label{eq:bootstrapstart}
    \frac{1}{8}\frac{1}{\DD_{12}+\#}(\DD_{23}^2D_{d,d+2,d+2,d}+\DD_{13}^2D_{d+2,d,d+2,d})\,.
\end{align}
While the relative coefficient is fixed by Bose symmetry, the overall coefficient is of course unfixed; here we made an arbitrary choice for convenience. 

We can then follow eq.~\eqref{eq:graviton_start} and~\eqref{eq:gr_recursion} to evaluate the action of $\frac{1}{\DD_{12}+\#}$. As already noted below eq.~\eqref{eq:gr_recursion}, requiring that the recursion truncates when $d$ is even leads to $\#=2d$. This is necessary for~\ref{r1} to hold. Thus, the above consideration leads us to
\begin{align}
    &\frac{1}{8}\frac{1}{\DD_{12}+2d}(\DD_{23}^2+\DD_{13}^2)D_{d,d,d,d}\nonumber\\
    &=2d^2(d+1)^2\left[P_{23}^2\frac{1}{\DD_{12}}D_{d,d+2,d+2,d}+P_{13}^2\frac{1}{\DD_{12}}D_{d+2,d,d+2,d}\right]-\frac{d^3(d-1)}{\DD_{12}-2d+4}D_{d,d,d,d}\nonumber\\
    &\quad +\frac{d^2}{2}D_{d,d,d,d}-\frac{d}{4}\frac{1}{\DD_{12}+2d}\DD_{12}D_{d,d,d,d}\,.
\end{align}
Although the recursion does truncate, the correlator still is not a finite sum of $D$-functions because of the last term in the above expression. The simplest solution is to add it to the left hand side. This indicates that the consistency with~\ref{r1} and~\ref{r2} naturally introduces additional interactions other than the one in~\ref{r3}.
It therefore follows that the ``bootstrap'' procedure described here yields the following s-channel contribution to the four-scalar correlator:
\begin{align}
    \mathcal{M}_{\rm bootstrap}(\text{s})=
    \frac{1}{8}\frac{1}{\DD_{12}+2d}(\DD^2_{23}+\DD^2_{13}+2d\DD_{12})D_{d,d,d,d}\,.
\end{align}
It differs from the true s-channel graviton exchange~\eqref{eq:grav_final} only by a contact term,
\begin{align}
    \hexchange{-0.5ex}{1}{2}{3}{4}&=\mathcal{M}_{\rm bootstrap}(\text{s})-\frac{1}{8}\DD_{12}D_{d,d,d,d}\,.
\end{align}
This additional contact term is exactly the contribution from the trace $h^{\nu}{}_{\nu}\nabla^{\mu}\varphi\nabla_{\mu}\varphi$. The difference is expected: our requirements cannot constrain polynomials of $\DD_{ij}$. Therefore, we have shown that based on three modest assumptions, we can reproduce the s-channel graviton exchange up to contact terms. If we further assume that the scalars do not have four-point contact interactions, then the only ambiguity is $\DD_{ij}D_{d,d,d,d}$. These terms sum up to zero if we add up all three exchange channels. As a result, under the additional assumption that scalar four-point interactions are absent, our bootstrap can reproduce the AdS boundary correlator of four identical scalars mediated by a graviton exchange,
\begin{align}
    \mathcal{M}(\varphi_1\varphi_2\varphi_3\varphi_4)=\mathcal{M}_{\rm bootstrap}(\text{s})+\mathcal{M}_{\rm bootstrap}(\text{t})+\mathcal{M}_{\rm bootstrap}(\text{u})\,.
\end{align}
It is not difficult to verify that the sum indeed agrees with eq.~\eqref{eq:AdSdc2} on the support of the conformal Ward identity, $(\DD_{12}+\DD_{23}+\DD_{13})D_{d,d,d,d}=0$.

\section{Conclusion \label{sec:conclusion}}

The differential representation of AdS boundary correlators is a new method for both the exploration of their properties and for their evaluation. By (conjecturally) organizing them solely in terms of generators of the AdS symmetry group (acting on a single contact integral), it has the ability to offer both a global perspective on all approaches to leading order AdS boundary correlators and to expose some of their hidden properties and unifying structures with flat space scattering amplitudes.
For example, it already provided a definition of the duality between color and kinematics and a link to an AdS generalization of the flat space BCJ amplitudes relations.

It is sometimes common in position-space formulations of QFTs to realize propagators of fields as nonlocal differential operators -- the inverse of the free-field operator. The curvature of space-time is an obstruction to the commutativity of these operators with, e.g., vertex factors. In this paper we showed that, in AdS space, the propagators can be written in terms of the quadratic Casimir operator of the AdS symmetry group and moreover that it can be made to act only on boundary data.  This organization is an essential step in the construction of differential representation of correlators, which posits a specific separation of the nonlocal (propagator) part of Witten diagrams from the local (vertex) part.\footnote{We remark that the separation that follows from the derivation in section~\ref{BGsection} applies equally well to bulk correlation functions. In close similarity with boundary correlators, the propagators become nonlocal differential operators acting only on the location of the bulk operator insertions.}
%
The correlators have a form remarkably similar to that of flat-space scattering amplitudes and we identified the appropriate replacements that yield the former from the latter.

A crucial ingredient in this separation and in the explicit examples we discussed is the AdS symmetry. From the perspective of the boundary theory, which is holographically-dual to the (gravitational or nongravitational) field theory in AdS space, this is captured by the conformal Ward identity. 
It plays a role analogous to flat-space momentum conservation, even more so than the relations between the Mellin-space analog of the Mandelstam variables. 

We demonstrated the construction of the differential representations at four and six points for scalar fields coupled to vectors, and four points when coupled to gravitons. The former exhibits color-kinematics duality and the AdS correlator relations. In both examples the local factors in each Witten diagram also bear close similarity with the kinematic numerator factors of the corresponding flat space amplitudes. 
We also prove that all scalar AdS boundary correlators have a differential representation of the type discussed in section~\ref{reviewdiffrep}.

While the differential representation of vector-field correlators remains elusive, the simplicity of scalar correlators together with the observation that scalar correlators can be understood as the dimensional reduction along boundary directions of vector-field correlators suggests that a differential representation for the latter may not be exceedingly involved. Furthermore, one can choose a particular gauge in the Berends-Giele recursion such that the resulting amplitudes are directly in the color-kinematics dual form~\cite{Mafra:2015vca,Mizera:2018jbh,Garozzo:2018uzj,Bridges:2019siz,Cheung:2021zvb}. It would be interesting to study if the Berends-Giele recursion presented in section~\ref{BGsection} can also be improved this way.

The differential representation exposes interesting and useful properties of AdS boundary correlators. To relate them to their position-space form we provide a general strategy that converts such a representation to a sums of contact integrals.
Using this method we recovered the classic result for the four-scalar correlator due to graviton exchange found in \cite{DHoker:1999kzh}.
Depending on the dimension of the boundary and on the mass parameters of the bulk fields (or, equivalently, the dimension of the boundary operators) the resulting sum may be finite or infinite. Interestingly, requiring that four-point correlators are given by a finite number of terms severely restricts their expression.\footnote{More specifically, it fixed their nonlocal parts up to arbitrary local terms.} 
This echoes the observation \cite{Rastelli:2017udc} such a truncation together with supersymmetry considerations provided very convenient means to uniquely fix four-point boundary correlators in maximal gauged supergravity in $\ads_5\times S^5$.

The construction described in this paper extends without difficulty to the other maximally-symmetric space, such as the de Sitter space (see \cite{Gomez:2021qfd,Gomez:2021ujt} for earlier work in this direction). Furthermore, a somewhat similar formalism was developed in Refs.~\cite{Baumann:2021fxj,Hillman:2021bnk} to study scale invariant theories in dS, where nested time integrals are traded for differential operators acting on flat space wavefunctions. 
More generally, it would be interesting to understand possible connections between the differential representation and other techniques for studying dS correlators, such as those in Refs. \cite{Arkani-Hamed:2015bza,Arkani-Hamed:2017fdk,Arkani-Hamed:2018kmz,Sleight:2019mgd,Hillman:2019wgh,Baumann:2020dch,Melville:2021lst,Goodhew:2021oqg}. 
Beyond dS, the formulation of the differential representation in terms of point-wise symmetry generators suggests that it may also be possible to extend our results to bulk correlation functions in more general symmetric spaces \cite{Binder:2020raz}. 

The double-copy construction in AdS and in more general curved spaces and its reliance on color-kinematics duality is one of the reasons for developing and studying the differential representation of AdS boundary correlators. While we have not discussed it at length in this paper, we comment here on several outstanding aspects.
We saw in section~\ref{gravexchange} that the flat space double-copy prescription, in which the color factor of one theory is replaced with the numerator factor of another, requires some modification. Indeed, part of the scalar correlator due to graviton exchange is related to scalar correlator due to gluon exchange. Additional numerator terms, proportional to the quadratic Casimir of the AdS symmetry group, are however necessary to produce the complete graviton exchange correlator. Further study is needed to understanding the systematics of these modifications.
The noncommutativity of the nonlocal and local factors in correlators suggests several possible double-copy prescriptions, especially for off-diagonal double-copies, which differ by the ordering of the Left and Right local and the nonlocal factors,
\begin{align}
    \hat{N}^{\text{L}}_{\rm s}\frac{1}{\DD_{12}}\hat{N}_{\rm s}^{\text{R}}\,,\quad 
    \hat{N}^{\text{R}}_{\rm s}\frac{1}{\DD_{12}}\hat{N}_{\rm s}^{\text{L}}\,,\quad 
    \frac{1}{\DD_{12}}\hat{N}^{\text{L}}_{\rm s}\hat{N}_{\rm s}^{\text{R}}\,,\quad\text{etc.}
    \label{prescriptions}
\end{align}
We have verified that a particular choice of prescription for the double copy of
the four-point NLSM  and gluon-mediated four-scalar correlators (scalar-YM) yields the local scalar correlator in the AdS Dirac-Born-Infeld theory (i.e. computed from the interaction $\mathcal{L}_{\rm int}\sim(\partial_{\mu}\phi\partial^{\mu}\phi)^2$): 
\begin{align}
\label{DBI}
    \mathcal{A}_{\text{DBI scalar}}&=(\DD_{12}^2+\DD_{23}^2+\DD_{13}^2)D_{d,d,d,d}\\
    &\sim\left[\hat{N}_{\rm s}^{\text{scalar-YM}}\frac{1}{\DD_{12}}\hat{N}^{\text{NLSM}}_{\rm s}+(\text{t- and u-channel})\right]D_{d,d,d,d}\,.\nonumber
\end{align}
The scalar-YM numerators are given in eq.~\eqref{eq:sYM}, and the NLSM numerators are given in eq.~(5.19) of ref.~\cite{Diwakar:2021juk}.
The other prescriptions in eq.~\eqref{prescriptions} do not fully cancel the nonlocal factor $\DD_{12}^{-1}$, suggesting that they belong to a theory that has nontrivial trilinear interactions. 
We note here that the double-copy prescription~\eqref{DBI} differs from the one relating the four-point scalar correlators mediated by gluon and graviton exchange discussed in section~\ref{gravexchange}. This suggests that more than one double-copy construction is possible.  
It would be important to understand their systematics and whether all possible prescriptions yield physical theories for a given pair of theories obeying color-kinematics duality.
As discussed in section~\ref{largeD}, for scalar boundary correlators the various conformal generators can be taken to commute in the large-$d$ limit when acting on a contact integral for states with $\Delta > d^{1/2}$ so all the prescriptions~\eqref{prescriptions} become identical with the flat space double-copy prescription in this limit.

The close relation between the nonlocal part of tree-level AdS boundary correlators and flat space tree-level amplitudes, given up to constant shifts by the replacement of Mandelstam invariants with the $SO(d,2)$ quadratic Casimir operators of the corresponding boundary insertion points, raises the question of the depth of their similarities. 
Being built on a Lagrangian, boundary correlators must exhibit some notion of unitarity and one may wonder if among its consequences is a form of multi-particle factorization at tree level. 
Inspection of the form of the differential representation suggests the analogs of the flat-space factorization poles corresponds to zeroes of differential operators $\DD_{i_1\dots i_k}+\mathsf{M}^2$. A careful definition may involve the decomposition of the boundary correlator in eigenfunctions of the such operators, identifying the corresponding residue as the part of the correlator that is in its kernel. It would be interesting to understand the relation between a differential formulation of factorization and others that have been discussed in the literature.\footnote{Factorization in Mellin space was discussed in \cite{Goncalves:2014rfa}, while position-space factorization corresponds to a decomposition of correlators in the possible intermediate states.}
If sufficiently precise, the notion of ``factorization on $\DD_{i_1\dots i_k}=-\mathsf{M}^2$ pole'' foreshadows a new formulation\footnote{Recursion relations for tree-level AdS boundary correlators in AdS momentum space were formulated in \cite{Raju:2011mp, Raju:2012zr}, using the split representation of bulk-bulk
propagators.} of on-shell recursion relations~\cite{Britto:2005fq} in AdS as well as a differential formulation\footnote{Formulations of unitarity-based methods in AdS were discussed e.g. in~\cite{Fitzpatrick:2011dm,Caron-Huot:2017vep,Simmons-Duffin:2017nub,Meltzer:2019nbs,Meltzer:2020qbr,Meltzer:2021bmb}. } of the generalized unitarity method~\cite{Bern:1994zx, Bern:1995db, Bern:1997sc, Bern:1994cg, Britto:2004nc}.
Together with a notion of integration over the space of conformal generators, along the lines of~\cite{Herderschee:2021jbi,Gomez:2021ujt}, it may lead to new approaches to loop-level AdS boundary correlators.

\acknowledgments
We would like to thank Clifford Cheung, Murat G\"unaydin, Song He, Aaron Hillman, Henrik Johansson, Arthur Lipstein, Paul McFadden, Jiajie Mei, Sebastian Mizera, Julio Parra-Martinez, Allic Sivaramakrishnan and Xinan Zhou for stimulating discussion. AH would like to especially thank Henriette Elvang for her continued support and comments. AH is supported in part by the US Department of Energy under Grant No. DE-SC0007859 and in part by a Leinweber Center for Theoretical Physics Graduate Fellowship. RR and FT are supported by the US Department of Energy under Grant No. DE-SC00019066.

\appendix

\section{Graviton bulk-bulk propagator in differential form}\label{sec:graviton}

In this appendix, we detail the derivation of eq.~\eqref{eq:bulkbulks}. To start with, we note that the graviton bulk-bulk propagator can be decomposed into a symmetric-traceless part ($\ell=2$) and a pure trace part ($\ell=0$) in the de Donder gauge~\cite{Costa:2014kfa,Sleight:2016hyl},
\begin{align}\label{eq:diffrep_step1}
    &\int_{\ads}dX_1dX_2T_{AB}(P_1,P_2,X_1) \prop_h^{AB,CD}(X_1,X_2) T_{CD}(P_3,P_4,X_2)\nonumber\\
    &=\int_{\ads}dX_1dX_2T(P_1,P_2,X_1,K_1)T(P_3,P_4,X_2,K_2)\prop_{(2)}(X_1,X_2,W_1,W_2)\nonumber\\
    &\quad +\int_{\ads}dX_1dX_2 \mathcal{T}(P_1,P_2,X_1)\mathcal{T}(P_3,P_4,X_2)\prop_{(0)}(X_1,X_2)\,.
\end{align}
where $\mathcal{T}(P_i,P_j,X)\equiv T_{AB}(P_i,P_j,X)G^{AB}$ is the trace of the vertex function. 
We have also used eq.~\eqref{eq:contraction} to express the index contraction in the $\ell=2$ part in terms of the operator $K_A$ and polarization $W_A$.

The $\ell=2$ part of the bulk-bulk propagator satisfy the equation~\cite{Costa:2014kfa,Sleight:2016hyl}
\begin{align}\label{eq:green}
    (\nabla_1^2 + 2) \prop_{(2)}(X_1,X_2,W_1,W_2) = -\delta(X_1,X_2) (W_1\cdot W_2)^2\, ,
\end{align}
with the Laplacian defined in eq.~\eqref{laplacian}.
Using eq.~\eqref{eq:green}, we can show that
\begin{align}
    &(\nabla_1^2 +2)\int_{\ads} dX_2 T(P_3,P_4,X_2,K_2) \prop_{(2)}(X_1,X_2,W_1,W_2)\nonumber\\
    &=-\int_{\ads} dX_2\delta(X_1,X_2)T(P_3,P_4,X_2,K_2) (W_1\cdot W_2)^2=-T(P_3,P_4,X_1,W_1)\,.
\end{align}
As a result, we can further transform eq.~\eqref{eq:diffrep_step1} into
\begin{align}
    &\int_{\ads}dX_1dX_2T(P_1,P_2,X_1,K_1)T(P_3,P_4,X_2,K_2)\prop_{(2)}(X_1,X_2,W_1,W_2)\nonumber\\
    &=-\int_{\ads}dX_1 T(P_1,P_2,X_1,K_1)\frac{1}{\nabla^2_1+2}T(P_3,P_4,X_1,W_1)\,.
\end{align}
It is also straightforward to show that
\begin{align}
    -(\nabla_1^2+2)T(P_3,P_4,X_1,W_1)=(\DD_{34}+2d)T(P_3,P_4,X_1,W_1)\,.
\end{align}
Therefore, we can trade $(\nabla^2_1+2)$ for $(\DD_{34}+2d)$ when acting on $T(P_3,P_4,X_1,W_1)$. This leads to the following expression for the $\ell=2$ exchange
\begin{align}\label{eq:diffrep_result}
     &\int_{\ads}dX_1dX_2T(P_1,P_2,X_1,K_1)T(P_3,P_4,X_2,K_2)\prop_{(2)}(X_1,X_2,W_1,W_2)\nonumber\\
     &=\frac{1}{\DD_{34}+2d}\int_{\ads}dX_1T(P_1,P_2,X_1,K_1)T(P_3,P_4,X_1,W_1)\nonumber\\
     &=\frac{1}{\DD_{34}+2d}\int_{\ads}dX T_{AB}(P_1,P_2,X)\Pi^{AB,CD}_{(2)}(X)T_{CD}(P_3,P_4,X)\,,
\end{align}
where $\Pi^{AB,CD}_{(2)}$ is the projector
\begin{align}
    \Pi^{AB,CD}_{(2)}(X)=\frac{1}{2}\left(G^{AC}G^{BD}+G^{AD}G^{BC}-\frac{2}{d+1}G^{AB}G^{CD}\right) \ .
\end{align}
To arrive at the last line of eq.~\eqref{eq:diffrep_result}, we have used the identity~\cite{Costa:2014kfa}
\begin{align}
    \frac{2}{(d-1)(d+1)}K_AK_B W^CW^D = \Pi^{AB,CD}_{(2)}\,.
\end{align}

The $\ell=0$ part of \eqref{eq:diffrep_step1} satisfies~\cite{Sleight:2016hyl}
\begin{align}
    (\nabla_1^2-2d)\prop_{(0)}(X_1,X_2)=\frac{2}{(d-1)(d+1)}\delta(X_1,X_2)
\end{align}
such that we can rewrite the last line of eq.~\eqref{eq:diffrep_step1} as
\begin{align}
\label{eq:traceexchange0}
    &\int_{\ads}dX_1dX_2 \mathcal{T}(P_1,P_2,X_1)\mathcal{T}(P_3,P_4,X_2)\prop_{(0)}(X_1,X_2)\nonumber\\
    &=\frac{2}{(d-1)(d+1)}\int_{\ads}dX_1 \mathcal{T}(P_1,P_2,X_1)\frac{1}{\nabla_1^2-2d}\mathcal{T}(P_3,P_4,X_1)\,.
\end{align}
Then, due to the identity
\begin{align}
    -(\nabla^2_1-2d)\mathcal{T}(P_3,P_4,X_1)=(\DD_{34}+2d)\mathcal{T}(P_3,P_4,X_1)\,,
\end{align}
the above equation becomes
\begin{align}
    &\int_{\ads}dX_1dX_2 \mathcal{T}(P_1,P_2,X_1)\mathcal{T}(P_3,P_4,X_2)\prop_{(0)}(X_1,X_2)\nonumber\\
    &=-\frac{2}{(d-1)(d+1)}\frac{1}{\DD_{34}+2d}\int_{\ads}dX  \mathcal{T}(P_1,P_2,X)\mathcal{T}(P_3,P_4,X) \\
    &=-\frac{2}{(d-1)(d+1)}\frac{1}{\DD_{34}+2d}\int_{\ads}dX  T_{AB}(P_1,P_2,X)G^{AB}G^{CD}T_{CD}(P_3,P_4,X)\,.\nonumber
\end{align}
Combining this with eq.~\eqref{eq:diffrep_result}, we get the following for the graviton exchange in terms of differential operators, 
\begin{align}
    &\int_{\ads}dX_1dX_2T_{AB}(P_1,P_2,X_1) \prop^{AB,CD}(X_1,X_2) T_{CD}(P_3,P_4,X_2)\\
    &=\frac{1}{\DD_{34}{+}2d}\frac{1}{2}\int_{\ads}dX T_{AB}(P_1,P_2,X)\Big[G^{AC}G^{BD}+G^{AD}G^{BC}-\frac{2}{d-1}G^{AB}G^{CD}\Big]\nonumber\\
    &\hspace{16em}\times T_{CD}(P_3,P_4,X)\,,\nonumber
\end{align}
which agrees with eq.~\eqref{eq:bulkbulks} since $\DD_{12}=\DD_{34}$ when acting on conformal partial waves.


\bibliographystyle{JHEP}
\bibliography{Draft.bib}

\end{document}